\documentclass{article}

\usepackage[english]{babel}
\usepackage[margin=1in, top=1.5in, bottom=1.5in, left=1.2in, right=1.2in]{geometry} 
\usepackage{graphicx,amsmath,amsfonts,amssymb,amsthm,float,bm,color,enumitem}
\usepackage[colorlinks=true,linkcolor=blue,citecolor=red]{hyperref}
\usepackage{subfigure}
\usepackage{authblk} 
\usepackage{sectsty} 

\newcommand{\email}[1]{\texttt{\small \href{mailto:#1}{#1}}} 
\newcommand{\affilnorm}[1]{\normalsize #1} 

\newtheorem{theorem}{Theorem} 
\newtheorem{lemma}{Lemma} 
\newtheorem{definition}{Definition} 

\usepackage{pifont}

\DeclareRobustCommand{\orcidicon}{%
    \begin{tikzpicture}
    \draw[lime, fill=lime] (0,0) 
    circle [radius=0.16] 
    node[white] {{\fontfamily{qag}\selectfont \tiny ID}};
    \draw[white, fill=white] (-0.0625,0.095) 
    circle [radius=0.007];
    \end{tikzpicture}
    \hspace{-2mm}
}

\newcommand{\orcidauthorA}{0000-0003-3611-0941}
\newcommand{\orcidauthorB}{0000-0003-2083-5032}
\newcommand{\orcidauthorC}{0000-0002-6917-3722}

\newcommand{\orcidA}{\href{https://orcid.org/\orcidauthorA}{\orcidicon}}
\newcommand{\orcidB}{\href{https://orcid.org/\orcidauthorB}{\orcidicon}}
\newcommand{\orcidC}{\href{https://orcid.org/\orcidauthorC}{\orcidicon}}

\usepackage{cite} 

\usepackage[figcolor=white]{todonotes}

\begin{document}

\title{\textsc{Heat and thermal travelling wave solutions of a nonlinear Maxwell-Cattaneo-Vernotte equation}}

\author[1,3]{C. F. Munaf\'o\orcidA\footnote{\email{carmelofilippo.munafo@unime.it}}}
\author[1]{P. Rogolino\orcidB\footnote{\email{progolino@unime.it}}}
\author[2]{M. Sciacca\orcidC\footnote{Co-corresponding author:\email{michele.sciacca@unipa.it}}}

\affil[1]{\affilnorm{Department of Mathematical and Computer Sciences, Physical Sciences and Earth Sciences, University of Messina, 98166, Messina, Italy}}
\affil[2]{\affilnorm{Engineering Department, University of Palermo, Viale delle Scienze, 90128, Palermo, Italy}} 
\affil[3]{\affilnorm{Department for the Promotion of Human Science and Quality of Life, San Raffaele Open University, Via di Val Cannuta, 247, 00166 Rome, Italy
}}
 
\date{}

\maketitle


\begin{abstract}
The propagation of heat and  thermal signals in the form of travelling waves is investigated for a nonlinear Maxwell–Cattaneo–Vernotte equation.
The  exact wave solutions are derived by expressing the thermal conductivity and the relaxation time as polynomial functions of  the temperature. This approach enables the identification of suitable degrees of nonlinearity that give rise to soliton solutions. Finally, exact solutions are shown through plots for the values of the selected parameters.
\end{abstract}


\section{\textsc{Introduction}} 
The linear theory of heat waves has played a key role in the development of non-equilibrium thermodynamics. Extending this framework to nonlinear regimes, particularly in the presence of sufficiently large amplitudes \cite{joseph1989heat,muller1998extended,jou2008EIT,jou2010EIT,rogolino2018generalized}, has therefore become a topic of significant interest. Although many nonlinear generalizations can be formulated, from a thermodynamic perspective it is crucial to identify those that are most directly compatible with the requirements imposed by the second law of thermodynamics, \cite{munafo2025comparison,ramos2024numerical}. 

Recently, soliton waves in heat transfer have been found by considering nonlinear lateral heat exchange with surroundings by radiative heat transfer\cite{sciacca2020heat,Sciacca_JNET47_2022nonlinear} and by flux-limited heat transport\cite{Sciacca_PD423_2021two,Sciacca_JMF62_2021thermal}, or by considering  a nonlinear Maxwell-Cattaneo–Vernotte (MCV) equation with special attention to high-amplitude wave propagation \cite{Sciacca_WM_2022thermal}. Here, instead, we use the relaxational Maxwell-Cattaneo–Vernotte (MCV) equation, assuming that both the thermal conductivity and the relaxation time depend nonlinearly on temperature.
This paper aims to find travelling thermal waves and solitons solutions as exact solutions of the mathematical model. From the application point of view, such solutions may provide efficient strategies for the transmission of thermal signals along thin one-dimensional devices, allowing propagation without dispersion or energy losses. This aspect is particularly relevant in the emerging field of phononics, where heat transport is exploited for information processing and control \cite{sciacca2020heat}.

\begin{figure}[!ht]
	\centering    
    \includegraphics[width=0.25\linewidth]{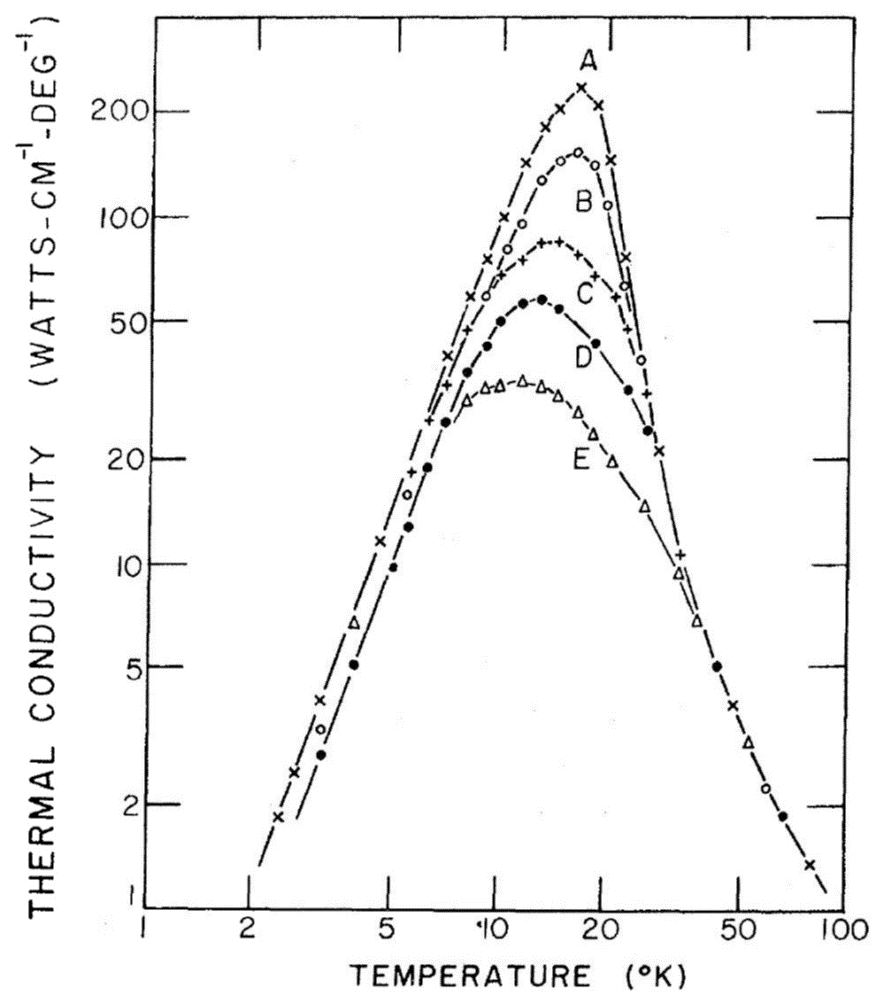}
	\caption{Plots of the thermal conductivity of NaF crystal with respect to the temperature for some values of the purity, from the highest A down to the lowest  E \cite{walker1971thermalconductivity}.}
	\label{McNelly_conductivity_NaF}	
\end{figure}

\begin{figure}[!ht]
	\centering
   	\includegraphics[width=0.7\textwidth]{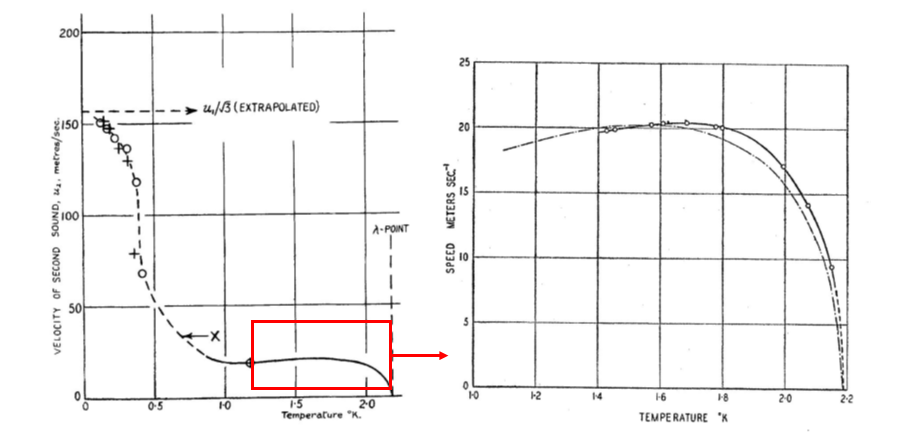}
	\caption{The temperature dependence of the propagation speed of second sound in He II \cite{atkins1950velocity,lane1947second}.}
	\label{second_sound_velocity}	
\end{figure}

The temperature dependence of thermal conductivity and relaxation time is of fundamental importance in the applications, particularly in  those systems where the propagation of the second sound (thermal waves) are used for studying some properties of the system, as for instance the density of quantized vortices in superfluid helium (He II) \cite{kapitza1941heat,dresner1984transient,Mongiovi_book2025non}, or the purity in NaF crystals. Figure \ref{McNelly_conductivity_NaF} provides  examples for various NaF crystals, showing the sensitivity of thermal conductivity to sample purity \cite{mcnelly1974PhD} as function of the temperature. 
In Fig. \ref{second_sound_velocity} it is shown how the velocity of the second sound depends on the temperature in He II in the range between $0\,K$ and $2.2\,K$ (the range of superfluidity), becoming  zero for temperature higher than the $\lambda$-point. This feature may also serve as an additional constraint to establish which temperature dependence is  physically admissible and to find out the relation between $\lambda(T)$ and $\tau(T)$.

The purpose of this paper is to  study  the propagation of thermal and heat waves in solid one-dimensional thin wires, and to check the conditions under which these waves arise. The mathematical model includes nonlinear terms arising from the temperature-dependence of the thermal conductivity $\lambda (T)$ and the relaxation time $\tau(T)$, once they are expanded into Taylor series  around the reference temperature. In particular,   the resulting wave solutions are analyzed as a function of the degree of polynomials that describe the thermal conductivity and the relaxation time. 
This approach allows us to identify specific degrees of nonlinearity that lead to the emergence of single-soliton solutions.

The paper is structured as follows. Section~\ref{sec2} introduces the mathematical model with particular emphasis on the nonlinear contributions of the relaxation time, thermal conductivity and mass density. In Section~\ref{sec3} the model is applied to a one-dimensional wire, and then it is dimensionless in order to facilitate the analysis of travelling wave solutions, which is carried out in Section~\ref{sec4}. More specifically,  Section~\ref{sec4} is for the analytical approach  used to find the  exact solutions of the governing equations. In section~\ref{sec5} these methods are applied to selected situations, including heat and thermal waves for the  model proposed in \cite{kovacs2020MCV1D}, the propagation of single-soliton solutions and train of solitons. Finally, Section~\ref{sec6} provides concluding remarks and outlines possible directions for future research.
\section{\textsc{The mathematical model}}\label{sec2}
In classical thermodynamics heat conduction is commonly described by the Fourier's law, which relates the heat flux $\mathbf q$ to the temperature gradient $\nabla T$ through the thermal conductivity $\lambda$. When this constitutive relation is combined with the internal energy balance equation
\begin{equation}\label{balance_energy}
    \rho\partial_t e + \nabla\cdot \mathbf q = 0, 
\end{equation}
assuming no internal heat sources and neglecting mechanical effects, the classical heat diffusion equation is obtained. In \eqref{balance_energy} $\rho$  denotes the mass density and $e$ the specific internal energy. Furthermore, $\partial_t (\star)$ denotes the partial time derivative, and $\nabla \cdot (\star)$ is the divergence operator.
However, Fourier's law implies an instantaneous response of the heat flux to the temperature gradient, leading to an infinite velocity of propagation of the temperature field. To overcome this limitation, the MCV equation introduces a finite relaxation time $\tau$ \cite{cattaneo1948sullaconduzione,toth2025unified}. The constitutive equation for heat conduction is therefore written as 
\begin{equation}\label{equation_CV}
    \tau \partial_t \mathbf{q} + \mathbf{q} = - \lambda \nabla T
\end{equation}
where $\tau$ is the relaxation time according for memory effects in heat transport.

The second law of thermodynamics provides a fundamental framework for deriving the Fourier heat equation as well as its generalizations, including the MCV equation. Thermodynamic admissibility of thermal processes is ensured by the requirement of non-negative entropy production, as imposed by the second law. In local form, this condition can be expressed as
\begin{equation}\label{2law}
    \sigma_s = \rho\, \partial_t s + \nabla \cdot \mathbf{J}_s \geq 0,
\end{equation}
where $s$ denotes the specific entropy and $\mathbf{J}_s = \frac{\mathbf q}{T}$ is the entropy flux vector. 
The heat flux is now treated as an additional state variable \cite{jou2008EIT}, so that the specific entropy depends on both the internal energy and the heat flux, namely $s = s(e,\mathbf q)$. 
The constitutive form  for the specific entropy is taken as
\begin{equation*}
	s(e,\mathbf{q}) = s_{eq}(e) - \frac{m(e)}{2} \mathbf{q} \cdot \mathbf{q},
\end{equation*}
where $s_{eq}(e)$ denotes the classical local equilibrium entropy and $m(e)$ is a positive function  representing  the coefficient of the non-equilibrium contribution. In general, the coefficient $m$ may depend on the internal energy $e$. Further details can be found in \cite{kovacs2020MCV1D}, where the the compatibility with the second law of thermodynamics was verified and, following the Onsager procedure, a nonlinear MCV heat equation was derived in which the thermal conductivity and the relaxation time are no longer constant but depend linearly on the temperature, as in \eqref{nonlinear1}. 

In particular, starting from the second law of thermodynamics
\begin{equation*}
    \Sigma_s = \rho \partial_t{s} + \nabla \cdot \mathbf{J}_s \geq 0
\end{equation*}
after some computations we get
\begin{equation}\label{entropy_production}
    \Sigma_s = \left[ \nabla\left( \frac{1}{T} \right) - \rho \, m(e) \partial_t \mathbf{q} - \left( \frac{1}{2} \frac{\mathrm{d} m(e)}{\mathrm{d}e} \nabla \cdot \mathbf{q} \right)\mathbf{q} \right] \cdot \mathbf{q}
\end{equation}
We restrict our analysis to the case in which $\frac{\mathrm{d} m(e)}{\mathrm{d} e}=0$, hence $m(e)$ is a constant, denoted by $\bar m$. Then, the entropy production \eqref{entropy_production} reduces to
\begin{equation*}
    \Sigma_s = \left[ \nabla\left( \frac{1}{T} \right) - \rho \,\, \bar m\,\,\partial_t \mathbf{q} \right] \cdot \mathbf{q}
\end{equation*}
In particular,  by applying  Onsager’s reciprocity framework \cite{Onsager1931reciprocal}, a linear relation between thermodynamic fluxes and forces can be established
\begin{equation*} 
    - \rho \bar m \partial_t \mathbf{q} + \nabla{\left(\frac{1}{T}\right)} = l \mathbf{q}, 
\end{equation*}
where $\bar{m}$ and $l$ are the positive phenomenological coefficients.
Introducing the identifications
\begin{equation*}
    \tau = \frac{\rho \bar m}{l}, \qquad \lambda = \frac{1}{l T^2} 
\end{equation*}
one obtains the nonlinear heat equation in the form
\begin{equation}
    \tau(T) \partial_t \mathbf{q} + \mathbf{q} = - \lambda(T)\nabla{T}. 
\end{equation}
Nonlinear extensions of the MCV equation and its coupling with the internal energy balance are important in heat transport, especially when temperature perturbations are not small compared to a reference temperature. In such cases, nonlinear effects arising from the temperature dependence of material properties, such as specific heat, thermal conductivity, and the relaxation time, become significant and play a non negligible role in the evolution of thermal disturbances.
 
In the papers \cite{kovacs2020MCV1D} and \cite{munafo2024nonlinear}, a generalization of the MCV heat equation, in one and two dimensions respectively, is introduced, in which both the thermal conductivity and the relaxation time are assumed to depend linearly on  temperature. Accordingly,
\begin{subequations}\label{nonlinear1}
    \begin{align}
        & \lambda(T)=\lambda_0+a(T-T_0), \label{nonlinear1.a} \\
        & \tau(T)=\tau_0+b(T-T_0), \label{nonlinear1.b}
    \end{align}
\end{subequations}
where $\lambda_0$ and $\tau_0$ denote the thermal conductivity and the relaxation time at the initial (or reference) temperature $T_0$, respectively. The coefficients $a$ and $b$ are material dependent parameters that may be either positive or negative and, when nonzero, introduce nonlinear behavior in the material properties.
Finally, we stress that both the thermal conductivity and the relaxation time must remain strictly positive, which imposes corresponding bounds on the admissible parameter values.

However, here we assume that the functions $\lambda(T)$ and $\tau(T)$ are arbitrary functions belonging to $\mathcal{C}^{\infty}$. Unlike the approach adopted by K\'ov\'acs and Rogolino in \cite{kovacs2020MCV1D}, where specific functional forms were prescribed (linear dependence on temperature), here we do not impose any particular structure on $\lambda(T)$ and $\tau(T)$. Since both functions are $\mathcal{C}^{\infty}$, we consider their Taylor's expansions around the reference temperature $T_0$,
\begin{align}
    \lambda (T) &= \sum_{i=0}^{+\infty} \, a_i \,\left( T-T_0 \right)^i \qquad \text{with} \qquad a_i = \left(\partial^i_{T^i}  \lambda\right)_{T=T_0}. \label{funzioni_lam} \\
    \tau (T) &= \sum_{j=0}^{+\infty} \, b_j \, \left( T-T_0 \right)^j \qquad \text{with} \qquad b_j = \left(\partial^j_{T^j} \tau\right)_{T=T_0}  \label{funzioni_tau}
\end{align}	
where $\partial^i_{T^i} f = \partial^i_{\underbrace{T,T,\dots,T}_{i-times}} f$ represents the partial derivative of the function $f$, $i-$times with respect to the variable $T$.

In the present analysis, for practical purposes, these series \eqref{funzioni_lam} and \eqref{funzioni_tau} are truncated at orders $n$ and $m$, respectively,
\begin{subequations}\label{tau_lambda}
	\begin{align}
    	\lambda (T) &= \sum_{i=0}^n \, a_i \,\left( T-T_0 \right)^i \label{tau_lambda.a} \\
        \tau (T) &= \sum_{j=0}^m \, b_j \, \left( T-T_0 \right)^j \label{tau_lambda.b}
    \end{align}	
\end{subequations} 
Nevertheless, the expansion can be extended to arbitrarily any order, in principle up to infinity, due to the regularity of the functions.
Hence,  the nonlinear generalized heat equation is expressed as follows:
\begin{equation}\label{MC_genre}
    \left( \sum\limits_{j=0}^m \, b_j \, \left( T-T_0 \right)^j \right) \partial_t \mathbf{q} + \mathbf{q} = - \left( \sum\limits_{i=0}^n \, a_i \,\left( T-T_0 \right)^i \right) \, \nabla{T}
\end{equation}

In order to obtain the temperature dependencies of $\lambda$ and $\tau$ given in \eqref{tau_lambda.a} and \eqref{tau_lambda.b}, the following constraints must be satisfied \cite{kovacs2020MCV1D}:
\begin{equation}\label{position_2}
    l(T) = \frac{1}{\left( \sum\limits_{i=0}^n \, a_i \,\left( T-T_0 \right)^i \right) T^2}, \quad
    \rho \bar m = \frac{\sum\limits_{j=0}^m \, b_j \, \left( T-T_0 \right)^j}{\left( \sum\limits_{i=0}^n \, a_i \,\left( T-T_0 \right)^i \right) T^2}.
\end{equation}
Under the assumption of a constant parameter $\bar{m}$, the above relations imply that the mass density $\rho$ must depend on the temperature $\rho(T)$, which is associated to the presence of mechanical effects. 
Alternatively, one could consider $\frac{d m(e)}{de}\neq 0$ which would introduce a new term into the Cattaneo equation, as seen in \cite{munafo2024PhD}. 

By exploiting \eqref{position_2}, the following expressions are obtained: 
\begin{equation}\label{rho_T}
    \rho(T) = \frac{b_0}{\bar m\,\left( \sum\limits_{i=0}^n \, a_i \,\left( T-T_0 \right)^i \right)\,T^2} + \frac{b_1(T-T_0)}{\bar m\,\left( \sum\limits_{i=0}^n \, a_i \,\left( T-T_0 \right)^i\right)\,T^2} + \dots + \frac{b_m (T-T_0)^m}{\bar m\,\left( \sum\limits_{i=0}^n \, a_i \,\left( T-T_0 \right)^i \right)\,T^2}
\end{equation}
namely,
\begin{equation*}
    \rho(T) = \sum_{j=0}^m \, \rho_j (T)
\end{equation*}
with 
\begin{equation}\label{rho_j}   
    \rho_j(T) = \frac{b_j(T-T_0)^j}{\bar m\,\left(\sum\limits_{i=0}^n \, a_i \,\left( T-T_0 \right)^i \right)\,T^2}, \quad j=0,1,\ldots m \dots.
\end{equation}
Then the mathematical model is expressed as
\begin{subequations}\label{MCV_3D}
	\begin{align}
    		& \left( \sum_{j=0}^m \, \rho_j(T) \right) c_v\, \partial_t T + \nabla \cdot \mathbf{q} = 0 \\
    		& \left( \sum_{j=0}^m \, b_j \, \left( T-T_0 \right)^j\right) \partial_t \mathbf{q} + \mathbf{q} = - \left( \sum_{i=0}^n \, a_i \,\left( T-T_0 \right)^i \right) \, \nabla{T}.
    \end{align}	
\end{subequations}
where the internal energy is expressed as $e=c_v T$, with $c_v$ the specific heat capacity.

\section{\textsc{One-dimensional formulation}}\label{sec3}
 Consider the set of equations \eqref{MCV_3D} in a one-dimensional domain $\Omega = [0,L]$ and $x$ the spatial coordinate
\begin{subequations}\label{MCV_1D}
	\begin{align}
    		& \left( \sum_{j=0}^m \, \rho_j(T) \right) c_v \partial_t T + \partial_x q = 0 \label{MCV_1D.1} \\
    		& \left( \sum_{j=0}^m \, b_j \, \left( T-T_0 \right)^j\right) \partial_t q + q = - \left( \sum_{i=0}^n \, a_i \,\left( T-T_0 \right)^i \right) \, \partial_x T \label{MCV_1D.2}
    \end{align}	
\end{subequations}
where $L$ denotes the length of the rigid and isotropic conductor.
The system \eqref{MCV_1D.1}-\eqref{MCV_1D.2} may involve coefficients spanning several orders of magnitude, which can lead to numerical difficulties. For this reason, it is advantageous to rescale the problem by introducing the following dimensionless variables:
\begin{equation*}
	\hat{x} = \frac{x}{L}, \qquad 
    \hat{t} = \frac{\alpha_{0} t}{L^2}, 
    \qquad \hat T = \frac{T - T_0}{T_{\mathrm{end}} - T_0}, \qquad
    \hat{q} = \frac{q}{\bar{q}_0}.
\end{equation*}
Here we also set 
\begin{equation*}
    \alpha_0 = \frac{a_0}{\rho_0\,c_v}, \qquad
    \rho_0  = \frac{b_0}{\bar m\,a_0\,T_0^2}, \qquad
     T_{{\mathrm{end}}} = T_0 +\dfrac{1}{\rho_0 c_v L} \int_{0}^{t_p} q_0(t)\; dt= T_0  +\frac{t_p \bar{q}_0}{\rho_0 c_v L},
\end{equation*}
where $a_0=\lambda(T_0)=\lambda_0$ from \eqref{tau_lambda.a},  $\rho_0= \rho (T_0)$ is the mass density value corresponding to the reference temperature $T_0$ and it comes from \eqref{position_2}, $t_p$ is the characteristic time and $\bar{q}_0$ represents the reference heat flux given by $\bar{q}_0=\frac{1}{t_p}\int_{0}^{t_p} q_0(t)\; dt$.

\begin{figure}[!ht]
	\centering
    	\includegraphics[width=0.65\linewidth]{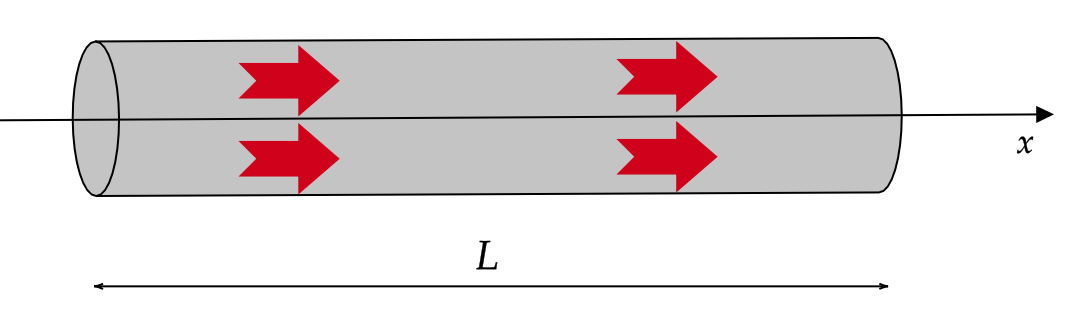}
	\caption{One dimensional domain $\Omega=[0,L]$}
	\label{1D_domain}	
\end{figure}
\noindent
We observe from \eqref{rho_j} that
\begin{equation*}
    \frac{\rho_j(T)}{\rho(T_0)} = \frac{b_j(T-T_0)^j}{b_0}, \quad j=0,1,\ldots, m
\end{equation*}
Assuming that 
\begin{equation*}
    \rho_0(T) \approx \rho(T_0)=\rho_0,
\end{equation*}
after many rearrangements, the dimensionless  equations \eqref{MCV_1D.1}-\eqref{MCV_1D.2} become
\begin{subequations}\label{dimensionless_MCV_1D}
	\begin{align}
  	     & \frac{\tau_{p_1}}{\tau_{q_1}} \left( \tau_{q_1} + \sum\limits_{j=1}^m\,\tau_{q_{j+1}} \hat{T}^j\right) \partial_{\hat{t}} \hat{T} + \partial_{\hat{x}} \hat{q} = 0, \label{dimensionless_MCV_1D.1}\\
         &\left( \tau_{q_1} + \sum\limits_{j=1}^m\,\tau_{q_{j+1}} \hat{T}^j \right) \partial_{\hat{t}} \hat{q} + \hat{q} = - \left(\tau_{p_1} + \sum\limits_{i=1}^n\,\tau_{p_{i+1}} \hat{T}^i \right)\partial_{\hat{x}} \hat{T},\label{dimensionless_MCV_1D.2}
	\end{align}
\end{subequations}	
where the dimensionless coefficients are defined as
\begin{align*}   
    & \tau_{p_1} = \frac{\alpha_0 t_p}{L^2}, \qquad
\tau_{p_{i+1}} = \frac{a_i(T_{\mathrm{end}}-T_0)\,t_p}{\rho_0 c_v L^2}, \qquad i=1,\dots,n; \\
    & \tau_{q_1} = \frac{\alpha_0 b_0}{L^2}, \qquad
\tau_{q_{j+1}} = \frac{\alpha_0, b_j\,(T_{\mathrm{end}}-T_0)}{L^2}, \qquad j=1,\dots,m \, .
\end{align*}
For further details, the reader is referred to Refs.~\cite{munafo2024PhD,kovacs2020MCV1D}.

Without loss of generality, the previous system \eqref{dimensionless_MCV_1D.1}-\eqref{dimensionless_MCV_1D.2}  can be written  
\begin{subequations}\label{sys_gen}
	\begin{align}
		& \alpha \left(\gamma_0 + \sum\limits_{j=1}^m \gamma_j U(x,t)^j \right) \,\partial_t U(x,t) + \partial_x V(x,t) = 0, \\
		& \left( \gamma_0 + \sum\limits_{j=1}^m \gamma_j U(x,t)^j \right) \partial_t V(x,t) + V(x,t) + \left( \beta_0 + \sum\limits_{i=0}^n \beta_i U(x,t)^i \right) \,\partial_x U(x,t) = 0,
	\end{align}
\end{subequations}	
where
\begin{equation*}
   U(x,t) = \hat{T}(x,t), \qquad V(x,t) = \hat{q}(x,t)
\end{equation*}
and the corresponding parameters are renamed  as
\begin{align*}
    \beta_0 = \tau_{p_1}, \qquad
    \beta_i = \tau_{p_{i+1}}, \qquad &i=1,\dots,n, \\
    \gamma_0 = \tau_{q_1}, \qquad
    \gamma_j = \tau_{q_{j+1}}, \qquad &j=1,\dots,m
\end{align*}
In the next section we provide a procedure for searching travelling waves solution of the   equations \eqref{sys_gen}. 

\section{\textsc{Travelling wave solutions}}\label{sec4}
The general solution of the partial differential equations \eqref{sys_gen} is an hard task to find, but it becomes easier in the context of travelling waves, especially in the moving frame of reference, because they become ordinary differential equations. Indeed, using the ansats $\xi=kx -wt$  ($k$ being the wavenumber of the wave and $w$ the frequency) the partial derivatives in  \eqref{sys_gen} becomes $\partial_t = -w \partial_{\xi}$ and $\partial_x = k \partial_{\xi}$, and  equations \eqref{sys_gen} change into  the following  ordinary differential equations
\begin{subequations}\label{sys1_gen}
	\begin{align}
    & - w \alpha \left( \gamma_0 + \sum_{j=1}^{m} \gamma_j \, U(\xi)^j \right) U'(\xi) + k \, V'(\xi) = 0 \label{sys1_gen_a}\\
    & - w \left( \gamma_0 + \sum_{j=1}^m \gamma_j \, U(\xi)^j \right) V'(\xi) + V(\xi) + k \left( \beta_0 + \sum_{i=0}^n \beta_i \, U(\xi)^i \right) U'(\xi) = 0 \label{sys1_gen_b}
    \end{align}
\end{subequations}
From  equation \eqref{sys1_gen_a} it is possible to obtain
\begin{equation*}
    V'(\xi) = \frac{w \alpha}{k} \left( \gamma_0 + \sum_{j=1}^{m} \gamma_j \, U(\xi)^j \right) \, U'(\xi)
\end{equation*}
and its solution is given by 
\begin{equation}\label{sol_V_gen}
	V(\xi) = \frac{w \alpha}{k} \left( \gamma_0 U(\xi) +  \sum\limits_{j=1}^m \frac{\gamma_j}{j+1} U(\xi)^{j+1} \right) + c_1 
\end{equation}
where $c_1 \in \mathbb{R}$ is an integration constant.

After substituting \eqref{sol_V_gen} into \eqref{sys1_gen_b} we obtain
\begin{multline*}
	- \frac{w^2 \alpha}{k} \left( \gamma_0 + \sum\limits_{j=1}^m \gamma_j U(\xi)^j \right)^2 \,U^{\prime}(\xi) + \frac{w \alpha}{k} \left( \gamma_0 U(\xi) + \sum\limits_{j=1}^m \frac{\gamma_j}{j+1} U(\xi)^{j+1} \right) + \\ +
    c_1 + k \left( \beta_0 + \sum\limits_{i=1}^n \beta_i U(\xi)^i \right) \, U^{\prime}(\xi) = 0 
\end{multline*}
from which
\begin{equation*}
    U^{\prime}(\xi) = \frac{w \alpha \left( \gamma_0 \, U(\xi) + \sum\limits_{j=1}^m \dfrac{\gamma_j}{j+1} U(\xi)^{j+1} \right) + k\,c_1}{w^2 \, \alpha \left( \gamma_0 + \sum\limits_{j=1}^m \gamma_j \, U(\xi)^j \right)^2 - k^2 \left( \beta_0 + \sum\limits_{i=1}^n \beta_i \, U(\xi)^i \right)}
\end{equation*}
and we get
\begin{equation}\label{integral_U}
	\int\dfrac{ w^2 \alpha \left( \gamma_0 + \sum\limits_{j=1}^m \gamma_j U(\xi)^j \right)^2 - k^2 \left( \beta_0 + \sum\limits_{i=1}^n \beta_i U(\xi)^i \right)}{w \alpha \left( \gamma_0 \, U(\xi) + \sum\limits_{j=1}^m \frac{\gamma_j}{j+1} U(\xi)^{j+1} \right) + k \, c_1 }\mathrm{d}U = \xi + c_2.
\end{equation}

Let's recall the following Hermite's Theorem.
\begin{lemma}[Hermite's Theorem]\label{Hermite_theorem}
    Let $P(x)$ and $Q(x)$ be two polynomial with real coefficients such that $\mathrm{deg}P<\mathrm{deg}Q$, and let
    \begin{equation*}
        Q(x) = a_n (x-b_1)^{n_1}\dots (x-b_j)^{n_j}
(x^2+c_1x+d_1)^{m_1}\dots (x^2+c_kx+d_k)^{m_k}
    \end{equation*}
    be the factorization of $Q(x)$ in $\mathbb{R}$.

    Then there exists the following unique expression for the function  $\frac{P(x)}{Q(x)}$ 
    \begin{equation*}
        \frac{P(x)}{Q(x)} = \frac{A_1}{x-b_1} + \dots + \frac{A_j}{x-b_j} + \frac{C_1 x + D_1}{x^2+c_1x+d_1} + \dots + \frac{C_k x + D_k}{x^2+c_kx+d_k} + \frac{d}{dx}\left(\frac{P^\star(x)}{Q^\star(x)}\right), 
    \end{equation*}
    where
    \begin{equation*}
        Q^\star(x) = (x-b_1)^{n_1-1}\dots (x-b_j)^{n_j-1}
(x^2+c_1x+d_1)^{m_1-1}\dots (x^2+c_kx+d_k)^{m_k-1},
    \end{equation*}
    and $P^\star(x)$ is a polynomial such that  
    \begin{equation*}
        \deg P^\star < \deg Q^\star.
    \end{equation*}
\end{lemma}
\begin{theorem}\label{integrable_theorem}
    The expression \eqref{integral_U} is integrable in closed form.
    \hfill $\qed$
\end{theorem}
Indeed, we can rewrite expression \eqref{integral_U} as
\begin{equation*}
	\int \frac{A(U)}{B(U)} \mathrm{d}U = \xi + c_2
\end{equation*}
with 
\begin{align*}
    \mathrm{deg}(B(U)) &= m+1\\ 
    \mathrm{deg}(A(U)) &= r = \max\{2m,n\} \geq m+1 = \mathrm{deg}(B(U)) \qquad  \forall n,m\neq 0
\end{align*}
The case $m=n=0$ is treated below. 

Being $\mathrm{deg}(A(U)) \geq \mathrm{deg}(B(U))$, except for $m=n=0$, we obtain
\begin{equation*}
	\int \frac{A(U)}{B(U)} \mathrm{d}U = \int Q(U) \, \mathrm{d}U + \int \frac{R(U)}{B(U)} \mathrm{d}U
\end{equation*}
where $Q(U)$ and $R(U)$ are the quotient and the rest of the division, with $\mathrm{deg}(R(U)) < \mathrm{deg}(B(U))$.

By applying Hermite’s theorem \eqref{Hermite_theorem} to the ratio $R(U)/B(U)$, we get
\begin{multline}\label{Hermite_appl}
	\int \frac{A(U)}{B(U)} \mathrm{d}U = \int Q(U) \, \mathrm{d}U + \int \frac{R(U)}{B(U)} \mathrm{d}U = \\
    = \underbrace{\int Q(U) \, \mathrm{d}U}_{1} + \underbrace{\int \left( \frac{A_1}{U-b_1} + \dots + \frac{A_N}{U-b_N} \right) \mathrm{d}U}_{2} + \\  \quad + \underbrace{\int \left( \frac{H_1 U + D_1}{U^2 + h_1 U + d_1} + \dots + \frac{H_M U + D_M}{U^2 + h_M U + d_M} \right) \mathrm{d}U}_{3} + \frac{R^\star(U)}{B^\star(U)}
\end{multline}
with $\mathrm{deg}(R^\star(U)) < \mathrm{deg}(B^\star(U))$ and the real roots $b_1<b_2<\ldots<b_N$ of $B(U)$. Thus, the integration of expression \eqref{integral_U} may be summarized as follows:

\begin{description}
    \item[Case $(m,n) = (0,0)$:] In this case expression                        \eqref{integral_U} becomes:
            \begin{equation*}
	           \int\dfrac{w^2 \alpha \gamma_0^2 - k^2 \beta_0}{w \alpha \gamma_0 \, U(\xi) + k \, c_1 }\mathrm{d}U = \xi + c_2
            \end{equation*}
            which can be integrated in 
            \begin{equation*}
                \dfrac{w^2 \alpha \gamma_0^2 - k^2 \beta_0}{w \alpha \gamma_0}\ \ln{|w \alpha \gamma_0 \, U(\xi) + k \, c_1 |} = \xi + c_2
            \end{equation*}
            and hence 
            \begin{equation*}
                |w \alpha \gamma_0 \, U(\xi) + k \, c_1 |= \exp{\left[\dfrac{w \alpha \gamma_0}{w^2 \alpha \gamma_0^2 - k^2 \beta_0} \left(\xi + c_2\right)\right]}
            \end{equation*}
        which leads to the function $U$ in terms of $\xi$.
    
    \item[Case $(m,n) \neq (0,0)$:] In this case,evaluating
        \eqref{integral_U} is related to the integral of the ratio $A[U]/B[U]$ as given in \eqref{Hermite_appl}, and hence to the expressions reported in the second and third lines of \eqref{Hermite_appl}. Below are the functions that can be obtained from these calculations:
        \begin{itemize}
            \item[\ding{202}] The integral of $Q(U)$ is expressed as
                \begin{equation*}
                    \int Q(U) \, \mathrm{d}U = \int \sum\limits_{i=0}^s \xi_i \, U^i \, \mathrm{d}U = \sum\limits_{i=0}^s \frac{\xi_i}{i+1} U^{i+1}
                \end{equation*}
                with $s=r-(m+1)$.
            \item[\ding{203}] Let assume that  $b_1<b_2<\ldots<b_N$. For two consecutive zeros, \emph{i.e.} $b_i$ and $b_{i+1}$, we set $y=U-\frac{b_i+b_{i+1}}{2}$ and we find:
                \begin{align*}
                    \frac{A_i}{U - b_i} + \frac{A_{i+1}}{U - b_{i+1}} &= \frac{A_i}{y+\frac{b_{i+1}}{2}-\frac{b_i}{2}} + \frac{A_{i+1}}{y+\frac{b_i}{2}-\frac{b_{i+1}}{2}} \\
                    & =\frac{(A_i + A_{i+1})y + (A_i - A_{i+1})\left( \frac{b_i}{2} - \frac{b_{i+1}}{2}\right)}{y^2 - \left( \frac{b_i}{2}-\frac{b_{i+1}}{2} \right)^2 }
                \end{align*}
                Thus, the above integral becomes 
                \begin{align}\label{asterisco}
                    \int \left( \frac{A_i}{U - b_i} + \frac{A_{i+1}}{U - b_{i+1}} \right) \mathrm{d}U & = \frac{A_i+A_{i+1}}{2} \ln\left|y^2 - \left( \frac{b_i}{2}+\frac{b_{i+1}}{2} \right)^2\right| \nonumber \\
                    & \quad + \dfrac{(A_i -A_{i+1})}{2} \ln{\left|\dfrac{y-(\frac{b_i}{2}-\frac{b_{i+1}}{2})}{y+(\frac{b_i}{2}-\frac{b_{i+1}}{2})}\right|}
                \end{align}
                which for $y/\left(\frac{b_i}{2}-\frac{b_{i+1}}{2}\right) \in (-1,1)$ becomes 
                \begin{align}\label{asterisco2}
                    \int \left( \frac{A_i}{U - b_i} + \frac{A_{i+1}}{U - b_{i+1}} \right) \mathrm{d}U &= \frac{A_i+A_{i+1}}{2} \ln\left|y^2 - \left( \frac{b_i}{2}+\frac{b_{i+1}}{2} \right)^2\right| \nonumber \\
                    & \quad - (A_i -A_{i+1}) \operatorname{artanh}\left(\frac{y}{\frac{b_i}{2}-\frac{b_{i+1}}{2}}\right),   
                \end{align}
                Note in \eqref{asterisco2} that for $A_1+A_{i+1}=0$ we find the inverse of the hyperbolic tangent function and for $A_1-A_{i+1}=0$ we find the logarithm function. 
                Finally,
                \begin{itemize}
                    \item if $N$ is even then we find that \ding{203} is given by $\frac{N}{2}$ terms like \eqref{asterisco} or \eqref{asterisco2};
                    \item if $N$ is odd, instead, \ding{203} is given by the sum of $\frac{N-1}{2}$ terms like  \eqref{asterisco} or \eqref{asterisco2}, and the last term is given by
                        \begin{equation*}
                            \int \frac{A_N}{U-b_N} \mathrm{d} U = A_N \log\left|U-b_N\right|
                        \end{equation*}
                \end{itemize}            
            \item[\ding{204}] The integration of the third term in \eqref{Hermite_appl} is given by
                \begin{align*}
                    \int\frac{H_j U + D_j}{U^2 + h_j U + d_j} \, d U & = 
                \frac{H_j}{2} \ln{(U^2 + h_j U + d_j)} \\ & \quad + \frac{2 D_j -H_j h_j}{\sqrt{4 d_j - h_j^2}} \arctan{\left( \frac{1}{\sqrt{4 d_j - h_j^2}} \left( U + \frac{h_1}{2} \right) \right)}
                \end{align*}
        \end{itemize}
\end{description}
Finally, the closed-form solution of the system \eqref{sys1_gen} is fulfilled whenever the functions $(U(\xi)$ and $V(\xi))$ can be explicitly expressed as functions of $\xi$. This is true for the function $V(\xi))$ through  the solution \eqref{sol_V_gen}. The function $U(\xi)$ is instead related to the inversion of  expression~\eqref{integral_U} once it is integrated  (which is guaranteed by the Theorem \ref{integrable_theorem}).

\section{\textsc{Thermal and heat waves}}\label{sec5}
In this section we focus our attention on some kind of travelling wave solutions which may be of interest in the one-dimensional physical systems, such as solid nanosystems. In the first subsection, we consider the propagation of thermal and heat waves of the mathematical model proposed in \cite{kovacs2020MCV1D}, whereas in the second and in the third subsection we focus our attention on the propagation of soliton waves. 

\subsection{Case $m=n=1$}\label{sec51}
This is the model considered by K\'ov\'acs and Rogolino in \cite{kovacs2020MCV1D}. In this case we obtain from \eqref{sol_V_gen}
\begin{equation}\label{solutionV_m1n1}
	V(\xi) = \frac{w \alpha}{k} \left( \gamma_0 \,U(\xi) + \frac{\gamma_1}{2} \,U(\xi)^2 \right) + c_1 
\end{equation}
where $c_1 \in \mathbb{R}$ is an integration constant, and from \eqref{integral_U}, after some computations we obtain
\begin{equation*}
	\int \frac{ w^2 \alpha \left( \gamma_0 + \gamma_1 U(\xi) \right)^2 - k^2 \left( \beta_0 + \beta_1 U(\xi) \right)}{w \alpha \left( \gamma_0 \, U(\xi) + \frac{\gamma_1}{2} U(\xi)^2 \right) + k \, c_1 }\mathrm{d}U = \xi + c_2.
\end{equation*}
Below we find two particular solutions: the travelling wave for $c_1 = \frac{w \alpha \gamma_0^2}{2 k \gamma_1}$ and the stationary solution for $w=0$.

\subsubsection{Travelling waves}
If we set $c_1=\frac{w \alpha \gamma_0^2}{2 k \gamma_1}$ we obtain
\begin{equation*}
	\int \frac{ 2 \gamma_1 w^2 \alpha \left( \gamma_0 + \gamma_1 U(\xi) \right)^2 - 2 \gamma_1 k^2 \left( \beta_0 + \beta_1 U(\xi) \right)}{w \alpha \left( \gamma_0 + \gamma_1 \,U(\xi) \right)^2 }\mathrm{d}U = \xi + c_2.
\end{equation*}
integrating yields
\begin{equation}\label{sol_U}
	2 \gamma_1 w \,U(\xi) - \frac{2 \gamma_1 k^2}{w \alpha} \left[ \frac{\beta_1}{\gamma_1^2} \ln(\gamma_0 + \gamma_1 \,U(\xi)) + \frac{\beta_0 \gamma_1 - \beta_1 \gamma_0}{\gamma_1^2 (\gamma_0 + \gamma_1 \,U(\xi))}\right] = \xi + c_2
\end{equation}
which cannot be inverted to explicitly find the function $U(\xi)$.

\subsubsection{Stationary waves}
The stationary solution is found by setting $w=0$ in \eqref{solutionV_m1n1} and in \eqref{sys1_gen_b}. We obtain the following expressions:
\begin{align*}
	&V^{\prime}(\xi) = 0 \\
	&V(\xi) + k \left( \beta_0 + \beta_1 \,U(\xi) \right) U^{\prime}(\xi) = 0
\end{align*}
from which we obtain $V(\xi) = c_3$ and 
\begin{equation*}
	c_3 + k \left( \beta_0 + \beta_1 \,U(\xi) \right) U^{\prime}(\xi) = 0.
\end{equation*}
We distinguish  two distinct cases:
\begin{itemize}
	\item the case $c_3 = 0$ leads to the following constant solutions
        \begin{equation*}
            U(\xi) = - \dfrac{\beta_0}{\beta_1} \qquad \text{or} \qquad U(\xi) = c_5
        \end{equation*}		 
	\item the case $c_3 \neq 0$ leads instead to the following solution  
        \begin{equation*}
            U(\xi) = \dfrac{-k \beta_0 \pm \sqrt{k^2 \beta_0^2 - 2 k \beta_1 (c_3 \xi + c_4)}}{k \beta_1}
        \end{equation*}	
    which is valid for $k^2 \beta_0^2 - 2 k \beta_1 (c_3 \xi + c_4) \geq 0$ or $2 k \beta_1 (c_3 \xi + c_4) \leq k^2 \beta_0^2$.
\end{itemize}

\subsection{Case $m=1$, $n=2$: single-soliton wave}\label{sec42}
In this case, the two expressions \eqref{sol_V_gen} and \eqref{integral_U} become 
\begin{equation*}
	V(\xi) = \frac{w \alpha}{k} \left( \gamma_0 \,U(\xi) + \frac{\gamma_1}{2} \,U(\xi)^2 \right) + c_1 
\end{equation*}
and
\begin{equation}\label{integral_m1n2}
	\int \frac{ w^2 \alpha \left(\gamma_0 + \gamma_1 \,U(\xi) \right)^2 - k^2 \left( \beta_0 + \beta_1 \,U(\xi) + \beta_2 \,U(\xi)^2 \right)}{ w \alpha \left( \gamma_0 \,U(\xi) + \frac{\gamma_1}{2} \,U(\xi)^2 \right) + k c_1 } \mbox{d}U = \xi + c_2.
\end{equation}
According to what found in  Section~\ref{sec3}, the general solution of \eqref{integral_m1n2} may be expressed in terms of polynomial, logarithms and/or the inverse function of trigonometric  or hyperbolic tangent, depending on the values of the parameters involved. 
In this subsection, we consider solutions which can be expressed in terms of hyperbolic tangent, which satisfies the definition of soliton, even though it is difficult to find a single, consensus definition of a soliton. In this paper we give the following definition: 

\begin{definition}\label{definition1}
    Let $u(x,t) = f(\xi = x-ct)$ be a function which satisfies a nonlinear partial differential equation $E(f,f_{\xi},..)= 0$. The function $f(\xi)$ is called \textsc{Soliton} if it is of permanent form, it is localized within a bounded domain and $\lim\limits_{\xi \to \infty} f(\xi)= y_0$, with $y_0$ a constant value.
\end{definition}
The kind of solitons which are usually found in literature are expressed in terms of $\operatorname{sech}(\xi)$ (in optics, bright solitons \cite{Agrawal_book2001, Hasegawa-book1995}) or $\tanh(\xi)$ (in optics and Bose-Einstein condensate, dark soliton \cite{kivshar1998dark, Frantzeskakis2010dark}). These functions satisfy the definition of soliton given in Definition~\ref{definition1}. 

In Section~\ref{sec3} we have seen that the integral \eqref{integral_m1n2} can be expressed in terms of the functions ``$\arctan(U)$'' and ``$\operatorname{artanh}(U)$'', which  lead to ``$\tan(\xi)$'' or ``$\tanh(\xi)$''.

For these reasons, we first make a change of variable and then we set two values for the parameters $\beta_1$ and $\beta_2$. 
Thus, by setting $y=U+ \frac{\gamma_0}{\gamma_1}$ in \eqref{integral_m1n2}, we obtain
\begin{equation*}
	\int \frac{w^2 \alpha \gamma_1^2 y^2 - k^2\beta_0 - k^2 \beta_1 y + k^2\beta_1 \frac{\gamma_0}{\gamma_1} - k^2 \beta_2 y^2 - k^2\beta_2 \frac{\gamma_0^2}{\gamma_1^2} + 2 k^2 \beta_2 \frac{\gamma_0}{\gamma_1} y}{\frac{w \alpha \gamma_1}{2 } y^2 - \frac{w \alpha \gamma_0^2}{2 \gamma_1} + c_1\,k} \mbox{d}y = \xi + c_2.
\end{equation*}
In the above expression, we assume that
\begin{align*}
	k^2 \beta_2 &= w^2 \alpha \gamma_1^2 \quad \Longrightarrow \quad \beta_2 = \frac{w^2 \alpha \gamma_1^2}{k^2}, \\
	k^2 \beta_1 &= 2 k^2 \beta_2 \frac{\gamma_0}{\gamma_1} \quad \Longrightarrow \quad  \beta_1 = 2 \beta_2 \frac{\gamma_0}{\gamma_1}=\frac{2 w^2 \alpha \gamma_1 \gamma_0}{k^2}
\end{align*}
and we get
\begin{equation}\label{int}
	2\frac{k^2 \beta_0 - w^2 \alpha \gamma_0^2}{w \alpha \gamma_1 \tilde{c}_1 } \int \frac{1}{1-\frac{1}{\tilde{c}_1} y^2} \mbox{d}y = \xi + c_2
\end{equation}
where $\tilde{c}_1 = \frac{2 k}{w \alpha \gamma_1} \left( \frac{w \alpha \gamma_0^2}{2 k \gamma_1} - c_1 \right) = \left( \frac{\gamma_0}{\gamma_1} \right)^2 - \frac{2 k c_1}{w \alpha \gamma_1}$. The integration of  \eqref{int} depends on the sign of $\tilde{c}_1$, in particular
\begin{equation}\label{split_functio}
    \int \frac{1}{1-\frac{1}{\tilde{c}_1} y^2} \mbox{d}y = 
        \begin{cases}
            \sqrt{\tilde{c}_1}\operatorname{artanh} \left( \dfrac{y}{\sqrt{\tilde{c}_1}} \right)  & \tilde{c}_1 > 0 \\[0.5cm]
            \sqrt{-\tilde{c}_1}\arctan \left( \dfrac{y}{\sqrt{-\tilde{c}_1}}\right) & \tilde{c}_1 < 0
        \end{cases}
\end{equation}
where for $\tilde{c}_1 > 0$ we have used the constraint $-1< \frac{y}{\sqrt{\tilde{c}_1}}<1$. Now, we distinguish  two cases: $\tilde{c}_1 > 0$ and  $\tilde{c}_1 < 0$.

\subsubsection{Case $\tilde{c}_1 > 0$ (Soliton wave)}

In this case, for $\tilde{c}_1 > 0$, the integral \eqref{int}  becomes 
\begin{equation}\label{artanh}
	\operatorname{artanh} \left( \frac{y}{\sqrt{\tilde{c}_1}} \right) = \frac{w \alpha \gamma_1 \sqrt{\tilde{c}_1} }{ 2 (k^2 \beta_0 - w^2 \alpha \gamma_0^2)} (\xi + c_2) .
\end{equation}
which leads to the solution
\begin{equation*}
    U(\xi) = \sqrt{\tilde{c}_1} \operatorname{tanh} \left( \frac{w \alpha \gamma_1 \sqrt{\tilde{c}_1} }{ 2 (k^2 \beta_0 - w^2 \alpha \gamma_0^2)} (\xi + c_2) \right) - \frac{\gamma_0}{\gamma_1} 
\end{equation*}

Thus, the solution of system \eqref{sys_gen} with $m=1$, $n=2$ is
\begin{subequations}\label{sol_ODE}
	\begin{align}
		U(\xi) &= \sqrt{\tilde{c}_1}\, \operatorname{tanh} \left( \frac{w \alpha \gamma_1 \sqrt{\tilde{c}_1} }{ 2 (k^2 \beta_0 - w^2 \alpha \gamma_0^2)} (\xi + c_2) \right) - \frac{\gamma_0}{\gamma_1} \label{sol_ODE_a}\\
		V(\xi) &= \left(c_1-\frac{w \alpha \gamma_0^2}{2 k \gamma_1}\right)  \operatorname{sech}^2\, \left( \frac{w \alpha \gamma_1 \sqrt{\tilde{c}_1} }{ 2 (k^2 \beta_0 - w^2 \alpha \gamma_0^2)} (\xi + c_2) \right)   \label{sol_ODE_b}
	\end{align}
\end{subequations}
If we set  
\begin{align*}
    A &=  \frac{w \alpha \gamma_1 \sqrt{\tilde{c}_1} }{ 2 (k^2 \beta_0 - w^2 \alpha \gamma_0^2} \\
    B &= \frac{w \alpha \gamma_1 \tilde{c}_1}{2 k}
\end{align*}
with $\tilde{c}_1 = \left( \frac{\gamma_0}{\gamma_1} \right)^2 - \frac{2 k c_1}{w \alpha \gamma_1}$, 
we get
\begin{subequations}\label{sol_ODE2}
	\begin{align}
		U(\xi) &= \sqrt{\tilde{c}_1}\, \operatorname{tanh} \left[A\, (\xi + c_2) \right] - \frac{\gamma_0}{\gamma_1} \label{sol_ODE_a2}\\
		V(\xi) &= -B\,  \operatorname{sech}^2 \left(A\, (\xi + c_2) \right) \label{sol_ODE_b2}
	\end{align}
\end{subequations}
The solution \eqref{sol_ODE2} in the one-dimensional system  becomes
\begin{subequations}\label{sol_ODE_final}
	\begin{align}
		u(x,t) &= \sqrt{\tilde{c}_1}\, \operatorname{tanh} \left[A\, (kx-wt + c_2) \right] - \frac{\gamma_0}{\gamma_1} \\
		v(x,t) &= -B\,  \operatorname{sech}^2 \left(A\, (kx-wt) + c_2) \right)
	\end{align}
\end{subequations}

Note that both the solutions in \eqref{sol_ODE_final} satisfy the definition of soliton given in Definition~\ref{definition1}. In particular, the wave $U(\xi)$ is a dark soliton (because $U(\xi)$ asymptotically goes to the constant value $\pm \sqrt{\tilde{c}_1} - \frac{\gamma_0}{\gamma_1}$ depending on the sign of $\infty$), instead  $V(\xi)$ is a bright soliton (because $V(\xi)$ asymptotically goes to zero).

The solution \eqref{sol_ODE2} for $c_2=0$ becomes
\begin{subequations}\label{sol_ODE3}
	\begin{align}
		U(\xi) &= \sqrt{\tilde{c}_1} \operatorname{tanh} \left[A\, \xi \right] - \frac{\gamma_0}{\gamma_1} \label{sol_ODE_a3}\\
		V(\xi) &= -B  \operatorname{sech}^2 \left[A\, \xi \right] \label{sol_ODE_b3}
	\end{align}
\end{subequations}

\begin{figure}[!ht]
    \centering
    \subfigure[]{\includegraphics[width=0.47\textwidth]{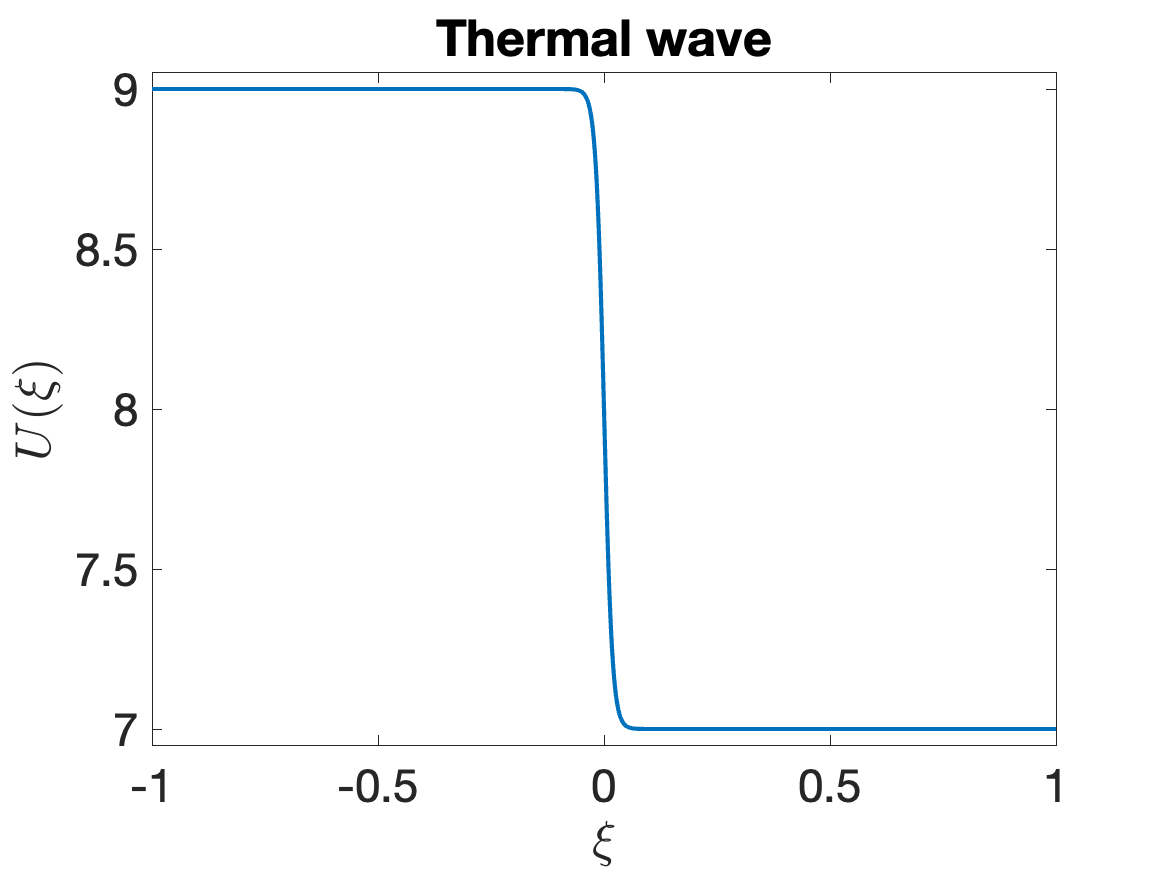}}
    \subfigure[]{\includegraphics[width=0.47\textwidth]{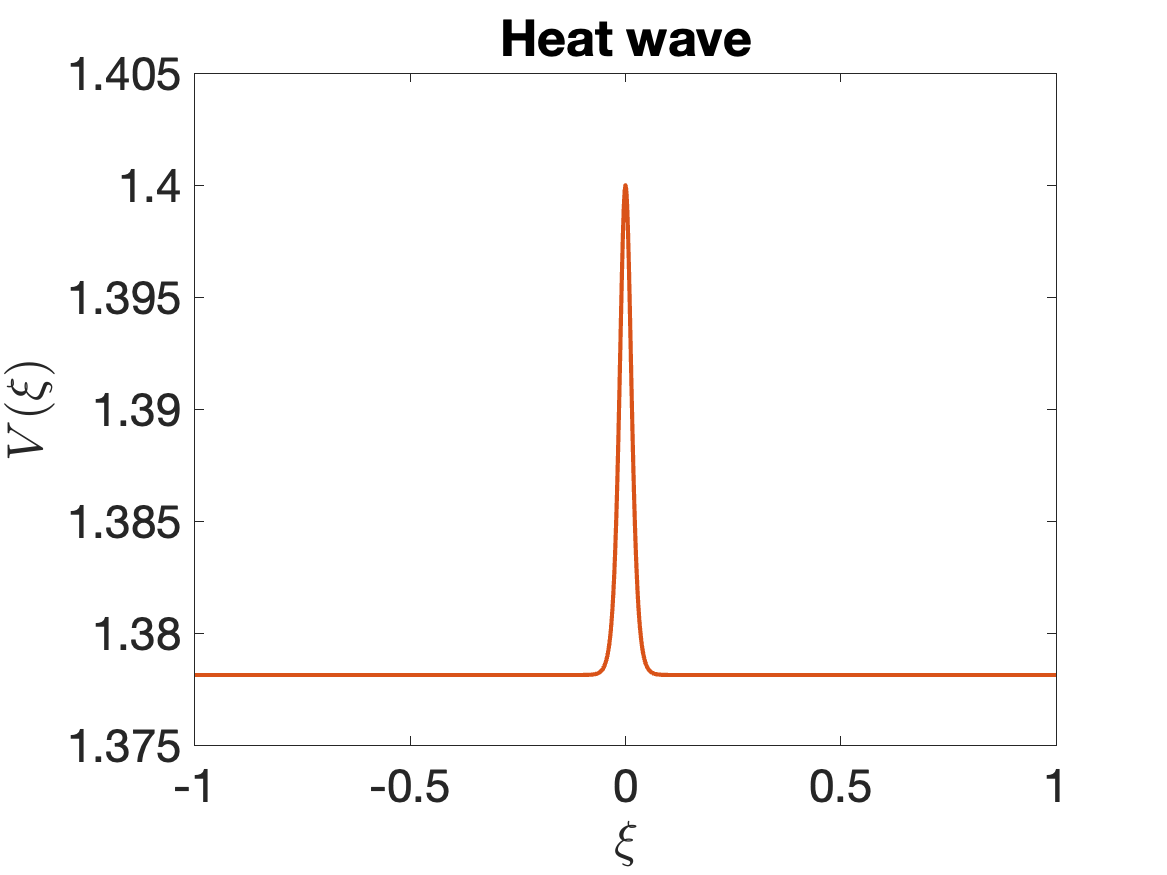}} \\
    \caption{Plot of the soliton solutions \eqref{sol_ODE}:  thermal dark soliton  $U(\xi)$ (left) and heat bright soliton $V(\xi)$ (right), with the following set of parameters: $A=1$, $\alpha=1.25$, $\gamma_0=0.08$, $\gamma_1=-0.01$, $\beta_0=0.1$, $k=0.2$, $w=0.7$, $c_1=0$ and $c_2=0$.}
    \label{fig:solution_ODE}
\end{figure}

\begin{figure}[!ht]
    \centering
    \subfigure[]{\includegraphics[width=0.47\textwidth]{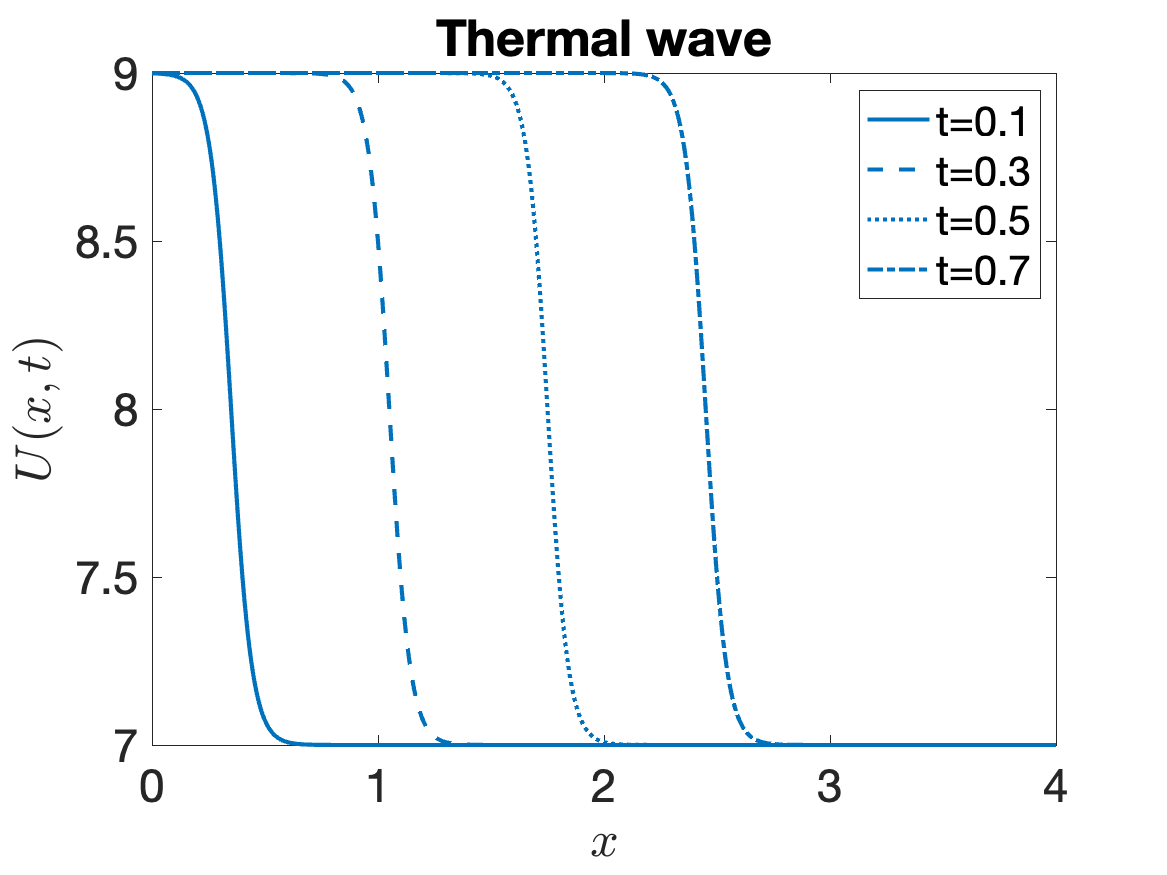}}
    \subfigure[]{\includegraphics[width=0.47\textwidth]{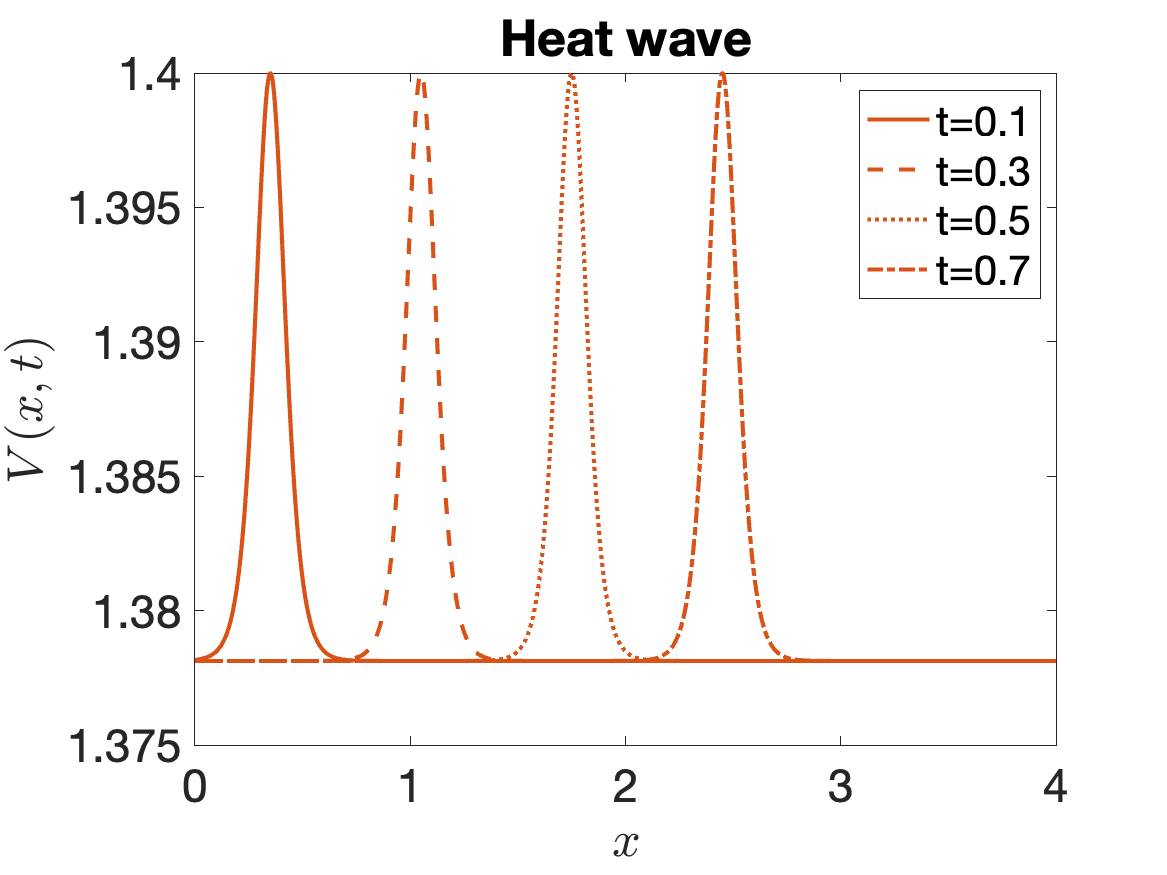}} \\
    \caption{Plot of the soliton solutions \eqref{sol_ODE} with the same parameters used in Fig.\ref{fig:solution_ODE} at different times.}
    \label{fig:solution_ODE}
\end{figure}

\begin{figure}[!ht]
    \centering
    \includegraphics[width=0.65\textwidth]{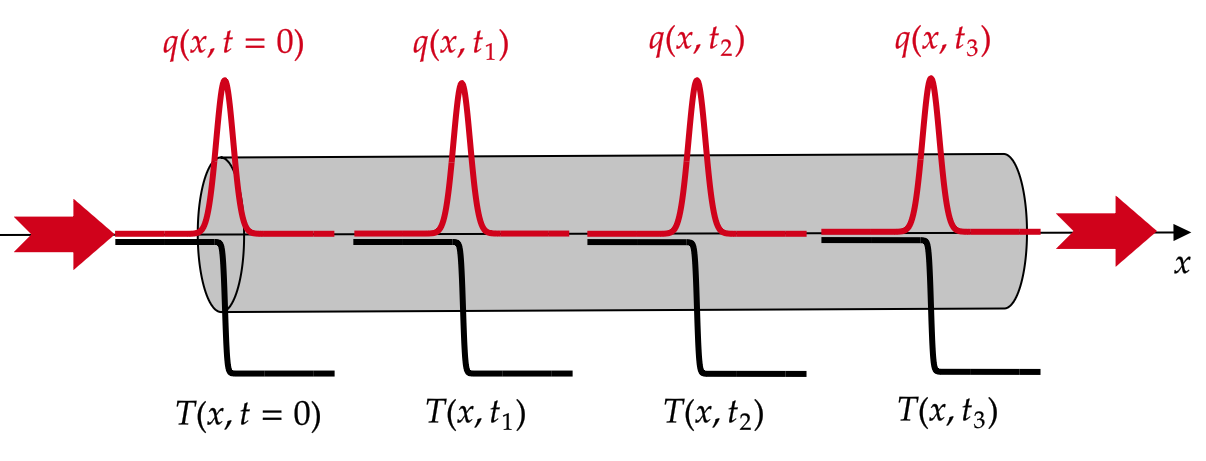}
    \caption{Sketch of two solitons, propagating along the cylinder: heat bright soliton waves (red profiles) and thermal dark soliton waves (black profile).}
    \label{fig:schema_soòution}
\end{figure}

\begin{figure}[!ht]
	\centering
    	\subfigure[]{\includegraphics[width=0.45\textwidth]{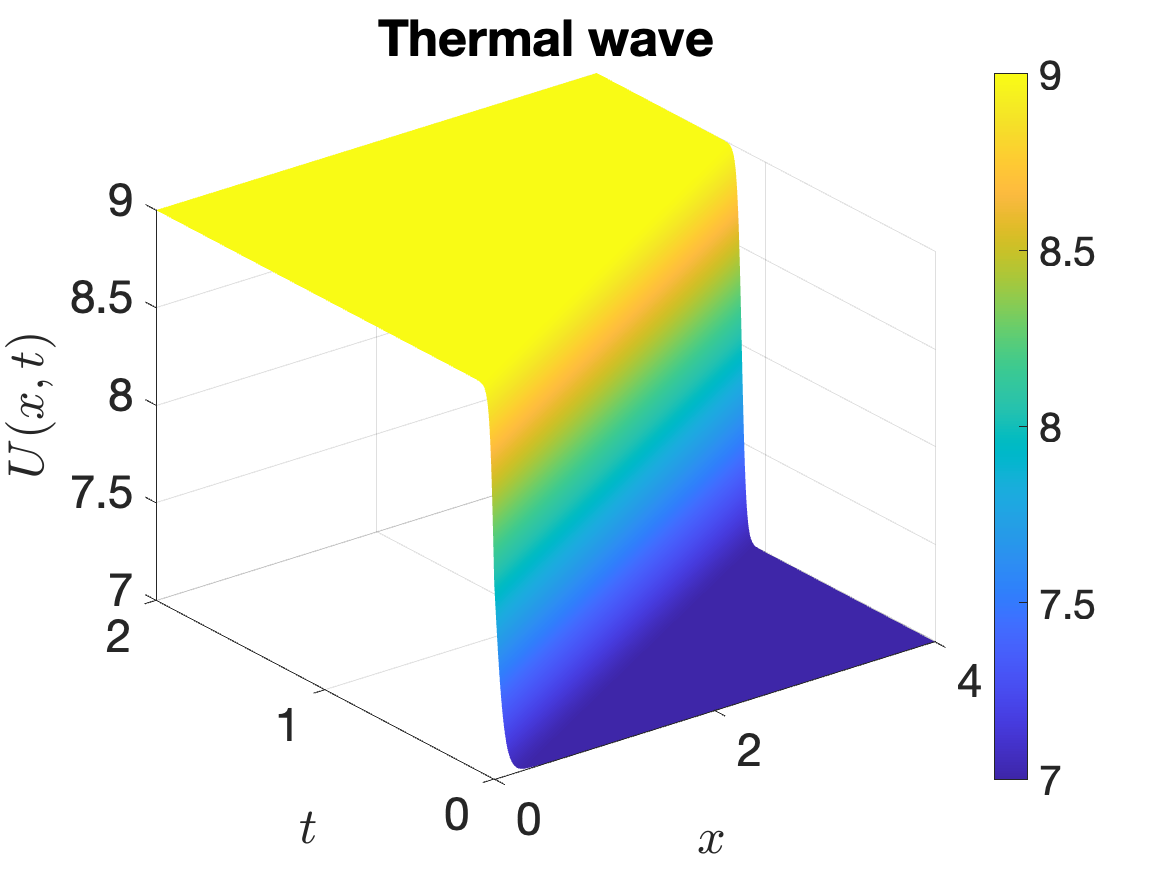}}
    	\subfigure[]{\includegraphics[width=0.45\textwidth]{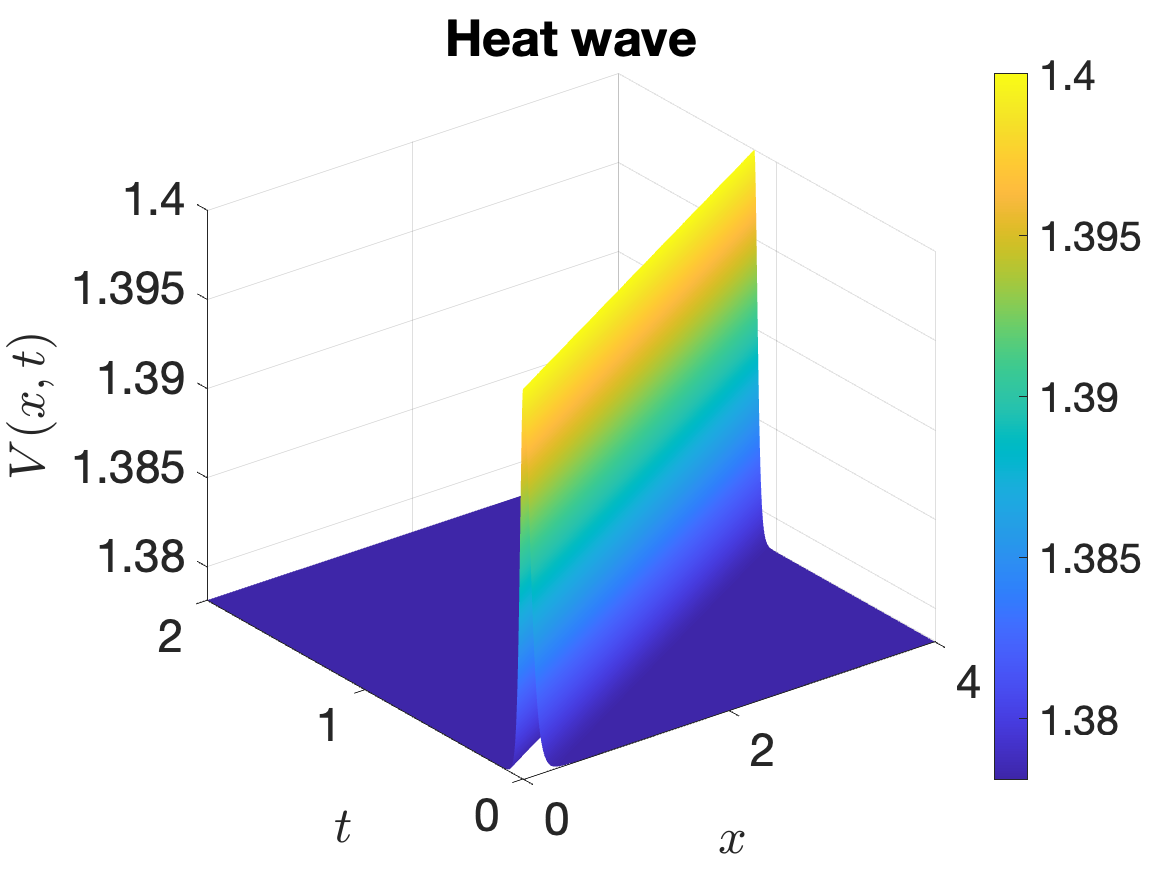}} \\
        \subfigure[]{\includegraphics[width=0.45\textwidth]{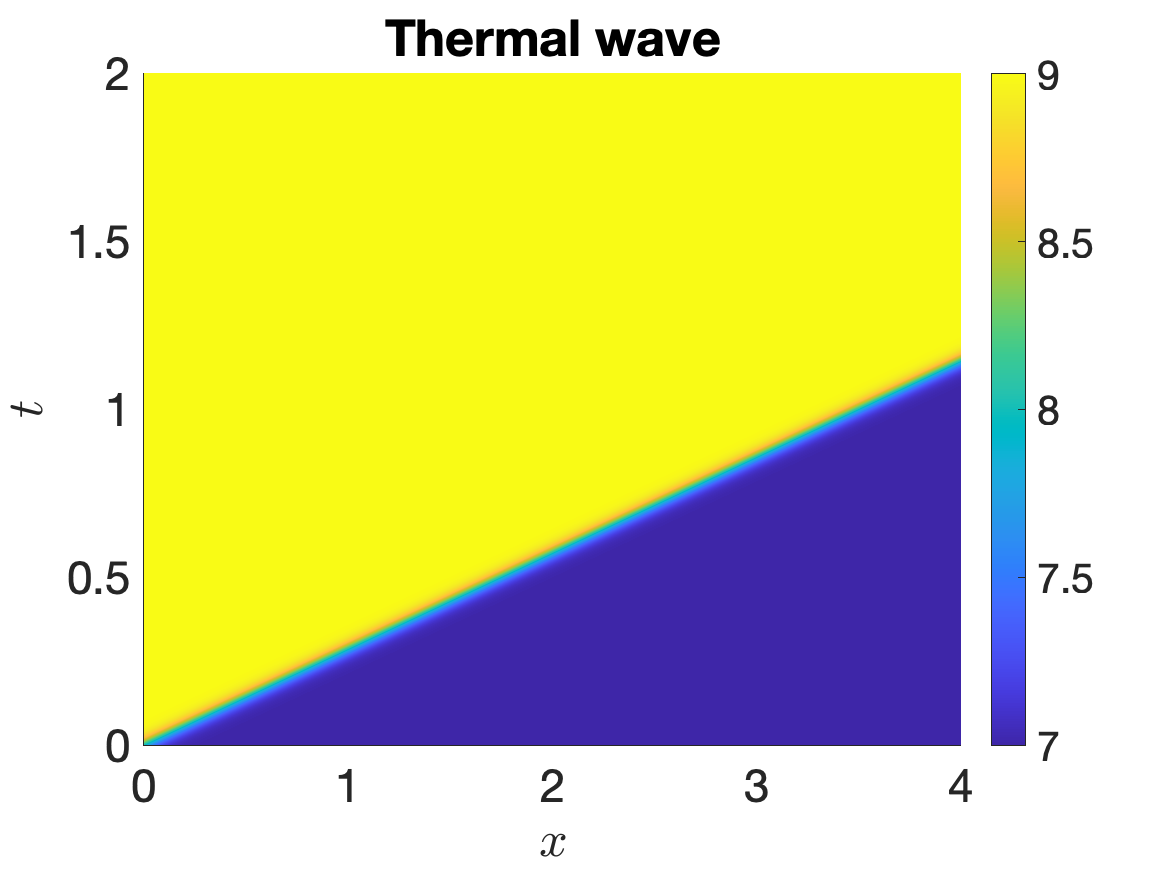}}
    	\subfigure[]{\includegraphics[width=0.45\textwidth]{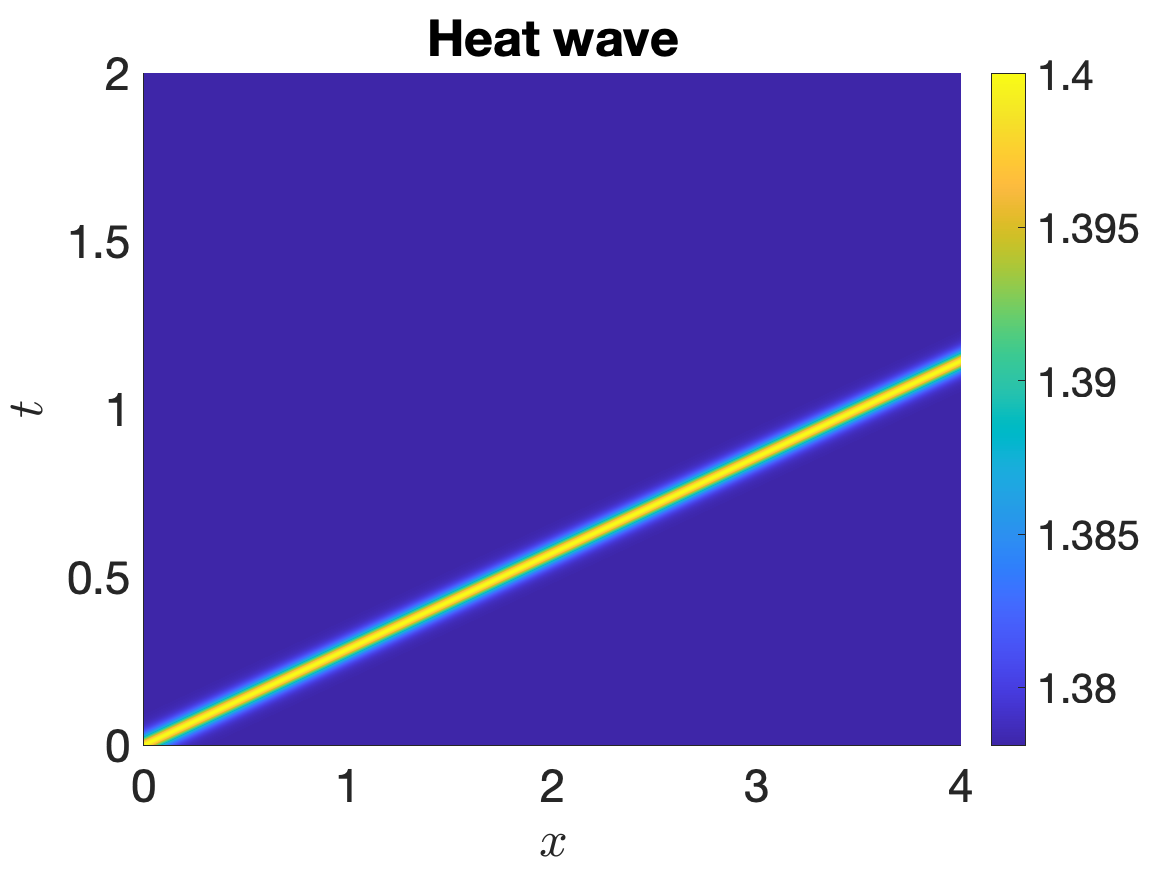}} \\
    	\caption{Plot of the soliton solutions \eqref{sol_ODE} in the $xt$-plane, namely $u(x,t)$ and $v(x,t)$, with $x \in [0,1]$ and $t\in\times[0,0.6]$, with the same choice of the parameters, namely: $A=1$, $\alpha=1.25$, $\gamma_0=0.08$, $\gamma_1=-0.01$, $\beta_0=0.1$, $k=0.2$, $w=0.7$, $c_1=0$ and $c_2=0$.}
    	\label{fig:solution_PDE}
\end{figure}
\subsubsection{Case $\tilde{c}_1 < 0$}

In this case, for $\tilde{c}_1 < 0$, the integral \eqref{int}  becomes 
\begin{equation*}
	\arctan \left( \frac{y}{\sqrt{-\tilde{c}_1}} \right) = -\frac{w \alpha \gamma_1 \sqrt{-\tilde{c}_1} }{ 2 (k^2 \beta_0 - w^2 \alpha \gamma_0^2)} (\xi + c_2) .
\end{equation*}
which leads to the solution
\begin{equation*}
    U(\xi) = \sqrt{-\tilde{c}_1} \tan \left( \frac{w \alpha \gamma_1 \sqrt{-\tilde{c}_1} }{ 2 (w^2 \alpha \gamma_0^2-k^2 \beta_0)} (\xi + c_2) \right) - \frac{\gamma_0}{\gamma_1} 
\end{equation*}

The solution of system \eqref{sys_gen} with $m=1$, $n=2$ is
\begin{subequations}\label{sol_ODEneg}
	\begin{align}
		U(\xi) &= \sqrt{-\tilde{c}_1} \tan \left( \frac{w \alpha \gamma_1 \sqrt{-\tilde{c}_1} }{ 2 (w^2 \alpha \gamma_0^2-k^2 \beta_0)} (\xi + c_2) \right) - \frac{\gamma_0}{\gamma_1}  \label{sol_ODE_nega}\\
		V(\xi) &= \frac{(-\tilde{c}_1) w \alpha \gamma_1 }{2 k}\, \sec^2\, \left( \frac{w \alpha \gamma_1 \sqrt{-\tilde{c}_1} }{ 2 (w^2 \alpha \gamma_0^2-k^2 \beta_0)} (\xi + c_2) \right) \label{sol_ODE_negb}
	\end{align}
\end{subequations}
If we set  
\begin{align*}
    A &=  \frac{w \alpha \gamma_1 \sqrt{-\tilde{c}_1} }{ 2 (w^2 \alpha \gamma_0^2-k^2 \beta_0)}  \\
    B &= \frac{(-\tilde{c}_1) w \alpha \gamma_1}{2 k}
\end{align*}
where $\tilde{c}_1 = \left( \frac{\gamma_0}{\gamma_1} \right)^2 - \frac{2 k c_1}{w \alpha \gamma_1}$, 
we get
\begin{subequations}\label{sol_ODE2neg}
	\begin{align}
		U(\xi) &= \sqrt{-\tilde{c}_1}\, \tan \left[A\, (\xi + c_2) \right] - \frac{\gamma_0}{\gamma_1} \label{sol_ODE_a2neg}\\
		V(\xi) &=  B\, \sec^2 \left(A\, (\xi + c_2) \right)  \label{sol_ODE_b2neg}
	\end{align}
\end{subequations}
The solution \eqref{sol_ODE2neg} in the cartesian variables becomes
\begin{subequations}\label{sol_ODE_finalneg}
	\begin{align}
		u(x,t) &= \sqrt{-\tilde{c}_1}\, \tan \left[A\, (kx-wt + c_2) \right] - \frac{\gamma_0}{\gamma_1} \\
		v(x,t) &= B\, \sec^2 \left(A\, (kx-wt + c_2) \right)
	\end{align}
\end{subequations}

The solution \eqref{sol_ODE2neg} for $c_2=0$ becomes
\begin{subequations}\label{sol_ODE3neg}
	\begin{align}
		U(\xi) &= \sqrt{-\tilde{c}_1}\, \tan \left[A\, \xi \right] - \frac{\gamma_0}{\gamma_1} \label{sol_ODE_a3neg}\\
		V(\xi) &=  B\, \sec^2 \left(A\, \xi \right)  \label{sol_ODE_b3neg}
	\end{align}
\end{subequations}
These functions do not satisfy the definition of soliton given in Definition~\ref{definition1}.

\subsection{Case $m=3$, $n=6$: train of soliton waves }\label{sec53}
The two polynomials in the numerator of the integral \eqref{integral_U} have the same power for $n = 2m$. In subsection~\ref{sec42} we have chosen $m=1$ and $n=2$ and we have found soliton waves (it was crucial the choice of $m$ for having two zeros in the denominator). In this subsection, we choose $m=3$ and $n=6$ so that the condition $n = 2m$ is satisfied which allows to counterbalance the two terms in the numerator, whereas $m=3$ leads to a fourth degree polynomial in the denominator for a possible appearance of two $\operatorname{artanh}$ as results of the integral. 

After setting $m=3$ and $n=6$ in \eqref{integral_U} we find
\begin{equation}\label{integral_U_sez43}
	\int\dfrac{ w^2 \alpha \left( \gamma_0 + \sum\limits_{j=1}^3 \gamma_j U(\xi)^j \right)^2 - k^2 \left( \beta_0 + \sum\limits_{i=1}^6 \beta_i U(\xi)^i \right)}{w \alpha \left( \gamma_0 \, U(\xi) + \sum\limits_{j=1}^m \frac{\gamma_j}{j+1} U(\xi)^{j+1} \right) + k \, c_1 }\mathrm{d}U = \xi + c_2,
\end{equation}
where we  assume that $c_1=0$, for the sake of simplicity, and that
\begin{equation*}
    \beta_4 = \frac{w^2 \alpha (\gamma_2^2 + 2 \gamma_1 \gamma_3)}{k^2}; \qquad \beta_5 = \frac{2 w^2 \alpha \gamma_2 \gamma_3}{k^2} \qquad \beta_6 = \frac{ w^2 \alpha \gamma_3^2}{k^2}; 
\end{equation*}
in order that the degree of the numerator is less than that of the denominator, and we get
\begin{equation*}
    \int\dfrac{ D_0 + D_1 U + D_2 U^2 + D_3 U^3}{w \alpha \left(\gamma_0 U + \frac{\gamma_1}{2} U^2 + \frac{\gamma_2}{3} U^3 + \frac{\gamma_3}{4} U^4 \right) }\mathrm{d}U = \xi + c_2,
\end{equation*}
with 
\begin{align*}
    D_0 &= w^2 \alpha \gamma_0^2 - k^2 \beta_0; \\
    D_1 &= 2 w^2 \alpha \gamma_0 \gamma_1 - k^2 \beta_1 \\
    D_2 &= w^2 \alpha (\gamma_1^2 + 2 \gamma_0 \gamma_2) -k^2 \beta_2 \\
    D_3 &= 2 w^2 \alpha (\gamma_1 \gamma_2 + \gamma_0 \gamma_3) -k^2 \beta_3.
\end{align*}

For the purpose of this subsection (namely, to obtain two $\operatorname{artanh}$ terms in the integral), we require that the denominator $D(U)$ has four real zeros. One of them is $U = 0$ (due to the assumption $c_1 = 0$ in \eqref{integral_U_sez43}), while the other three are denoted by $b_1$, $b_2$, and $b_3$. Thus, the denominator can be factorized as
\begin{equation*}
    D(U) = w \alpha \left(\gamma_0 U + \frac{\gamma_1}{2} U^2 + \frac{\gamma_2}{3} U^3 + \frac{\gamma_3}{4} U^4 \right) = w \alpha b_0 \, U (U - b_1) (U - b_2) (U - b_3).
\end{equation*}
where $b_1$, $b_2$, and $b_3$ can be determined by equating the two polynomials.
In particular, to achieve this decomposition, the following constraints must be satisfied by the coefficients
\begin{equation*}
    \gamma_0 = - c_0 b_1 b_2 b_3, \qquad \gamma_1 = 2 c_0 ( b_1 b_2 + b_1 b_3 + b_2 b_3), \qquad \gamma_2 = - 3 c_0 (b_1 + b_2 + b_3),
\end{equation*}
with $c_0 = \dfrac{\gamma_3}{4}$. We also require that, after the translation 
\begin{equation*}
    y = U - \frac{b_1}{2}, 
\end{equation*}
the zeros of the polynomial $D(y)$ are symmetric with respect to the origin, but the zeros of
\begin{equation*}
    D(y) = w \alpha c_0 \, \left( y + \frac{b_1}{2} \right) \left( y - \frac{b_1}{2} \right) \left( y + \left(\frac{b_1}{2} - b_2 \right) \right) \left( y + \left(\frac{b_1}{2} - b_3 \right) \right)
\end{equation*}
are symmetric if the following constraint $\left(\frac{b_1}{2}-b_2\right)=-\left(\frac{b_1}{2}-b_3\right)$, \emph{i.e.} $b_3=b_1-b_2$, is verified, given that the first two zeros are symmetric by construction. This constraint simplifies the denominator into
\begin{equation*}
    D(y) = w \alpha c_0\, \left( y^2 - \left( \frac{b_1}{2} \right)^2 \right) \left( y - \left(\frac{b_1}{2} -b_2 \right)^2 \right)
\end{equation*}
Thus, the integral \eqref{integral_U_sez43} becomes
\begin{equation}\label{integral_U_sez43_2}
	\frac{1}{w \alpha c_0} \int \dfrac{A_0}{y-\frac{b_1}{2}} + \dfrac{A_1}{y+\frac{b_1}{2}}+\dfrac{A_2}{y-\left(\frac{b_1}{2}-b_2\right)}+\dfrac{A_3}{y+\left(\frac{b_1}{2}-b_2\right)}\mathrm{d}y = \xi + c_2.
\end{equation}
where
\begin{align*}
        A_0 &= \frac{1}{16 B_1 (B_1^2-B_2^2)} \left[ k^2( 8 \beta_0 - 4 (b_1 -2 B_1) \beta_1 + 2 (b_1 - 2 B_1)^2 \beta_2 - (b_1 - 2 B_1)^3 \beta_3) \right. \\ & \qquad \left. + 2 w^2 \alpha ( -2 \gamma_0 + (b_1 - 2 B_1) \gamma_1) (2 \gamma_0 + (b_1 - 2 B_1)( - \gamma_1 + (b_1 - 2 B_1) \gamma_2)) + (b_1 - 2 B_1)^3 \gamma_0 \gamma_3) ) \right]; \\ \\
        A_1 &= \frac{1}{16 B_1 (B_1^2-B_2^2)} \left[ k^2( -8 \beta_0 + 4 (b_1 + 2 B_1) \beta_1 - 2 (b_1 + 2 B_1)^2 \beta_2 + (b_1 + 2 B_1)^3 \beta_3) \right. \\ & \qquad \left. + 2 w^2 \alpha ( -2 \gamma_0 + (b_1 + 2 B_1) \gamma_1) (-2 \gamma_0 + (b_1 + 2 B_1) (\gamma_1 - (b_1 + 2 B_1) \gamma_2) ) - (b_1 + 2 B_1)^3 \gamma_0 \gamma_3)) \right]; \\ \\
        A_2 &= \frac{1}{16 B_2 (B_2^2-B_1^2)} \left[ k^2( 8 \beta_0 - 4 (b_1 - 2 B_2) \beta_1 + 2 (b_1 - 2 B_2)^2 \beta_2 - (b_1 - 2 B_2)^3 \beta_3) \right. \\ & \qquad \left. + 2 w^2 \alpha ( -2 \gamma_0 + (b_1 - 2 B_2) \gamma_1) (2 \gamma_0 + (b_1 - 2 B_2) (- \gamma_1 + (b_1 - 2 B_2) \gamma_2)) + (b_1 - 2 B_2)^3 \gamma_0 \gamma_3) ) \right]; \\ \\
        A_3 &= \frac{1}{16 B_2 (B_2^2-B_1^2)} \left[ k^2( - 8 \beta_0 + 4 (b_1 + 2 B_2) \beta_1 - 2 (b_1 + 2 B_2)^2 \beta_2 + (b_1 + 2 B_2)^3 \beta_3) \right. \\ & \qquad \left. + 2 w^2 \alpha ( -2 \gamma_0 + (b_1 + 2 B_2) \gamma_1) (-2 \gamma_0 +(b_1 + 2 B_2) (\gamma_1 - (b_1 + 2 B_2) \gamma_2)) - (b_1 + 2 B_2)^3 \gamma_0 \gamma_3) ) \right]; 
\end{align*}
with 
\begin{equation*}
    B_1 = \frac{b_1}{2}, \qquad B_2 = \frac{b_1}{2} - b_2
\end{equation*}
If we further impose the conditions $A_0+A_1=0$ and $A_2+A_3=0$, then the integral \eqref{integral_U_sez43_2} can be written
\begin{equation}\label{integral_U_sez43_3}
	-\dfrac{1}{w\alpha c_0}\int \frac{2 A_0 B_1}{\left(B_1^2- y^2 \right)} + \frac{2 A_2 B_2}{\left(B_2^2 - y^2\right)}
    \mathrm{d}y = \xi + c_2.
\end{equation}
This expression \eqref{integral_U_sez43_3} can then be integrated to obtain
\begin{equation}\label{integral_U_sez43_4}
	- \frac{2 A_0}{c_0 w \alpha} \operatorname{artanh} \left( \frac{2y}{b_1}\right) - \frac{2 A_2}{c_0 w \alpha} \operatorname{artanh} \left( \frac{2y}{b_1 - 2 b_2} \right) = \xi + c_2
\end{equation}
restricted in the range $-y_L < y < y_L$ with $y_L = \min \left\{\dfrac{|b_1|}{2}, \dfrac{|b_1-2 b_2|}{2} \right\}$. 
The constraints $A_0+A_1=0$, $A_2+A_3=0$ and the $A_0=A_2$ mean
\begin{align*}
    \beta_1 &= - \frac{ 2 c_0^2 b_1 b_2 w^2 \alpha (3 b_1 - 2 b_2) (b_1^2 - b_2^2)}{k^2} + \frac{2 \beta_0}{b_2} \\
    \beta_2 &= \frac{ 2 c_0^2 w^2 \alpha (2 b_1^4 + 9 b_1^3 b_2 - 6 b_1^2 b_2^2 - 5 b_1 b_2^3 + 2 b_2^4)}{k^2} + \frac{2 \beta_0}{b_1 b_2} \\
    \beta_3 &= \frac{8 c_0^2 b_1 w^2 \alpha (3 b_1 - 2 b_2) (b_1 + 2 b_2)}{k^2}
\end{align*}

The solution of \eqref{integral_U_sez43_4} is analogous to that of \eqref{artanh}. In the general case, with two hyperbolic arctangent terms having distinct arguments, the solution can be interpreted as the superposition of two solitons.
Using relation $\operatorname{artanh}(X)=\frac{1}{2} \ln \left(\frac{1+X}{1-X} \right)$ equation \eqref{integral_U_sez43_4} becomes 
\begin{equation}\label{integral_U_sez43_4bis}
	\left( \frac{b_1 + 2y}{b_1 - 2y} \right)^{\frac{A_0}{c_0 w \alpha}}
    \left( \frac{b_1 - 2b_2 + 2y}{b_1 - 2b_2 - 2y}\right)^{\frac{A_2}{c_0 w \alpha}} = \operatorname{e}^{-c_2} \operatorname{e}^{-\xi}  
\end{equation}
The inversion of the solution \eqref{integral_U_sez43_4} depends on the values of the coefficients $A_0$ and $A_2$. For instance, if we choose $A_0=A_2$ (as previously  mentioned), the above equation leads to the solution
\begin{equation*}
    y(\xi) = \frac{- 2b_1 + b_2 \left(1+ \operatorname{e}^{\frac{c_0 w \alpha}{A_0} (\xi+c_2)}\right) \pm \sqrt{ 2 \operatorname{e}^{\frac{c_0 w \alpha}{A_0} (\xi+c_2)} + \left(2 b_1^2 - 4 b_1 b_2 +b_2^2 + b_2^2 \cosh{\left(\frac{c_0 w \alpha}{A_0} (\xi + c_2)\right)} \right) }}{2 \left( \operatorname{e}^{\frac{c_0 w \alpha}{A_0} (\xi+c_2)} - 1 \right)}
\end{equation*}
then the solution is 
\begin{equation}\label{solU_case_n_6_m_3}
    \begin{aligned}
        U(\xi) &= -\frac{b_1}{2} + \frac{- 2b_1 + b_2 \left(1+ \operatorname{e}^{\frac{c_0 w \alpha}{A_0} (\xi+c_2)}\right) \pm \sqrt{ 2 \operatorname{e}^{\frac{c_0 w \alpha}{A_0} (\xi+c_2)} + \left(2 b_1^2 - 4 b_1 b_2 +b_2^2 + b_2^2 \cosh{\left(\frac{c_0 w \alpha}{A_0} (\xi + c_2)\right)} \right) }}{2 \left( \operatorname{e}^{\frac{c_0 w \alpha}{A_0} (\xi+c_2)} - 1 \right)} \\
        V(\xi) &= \frac{w \alpha}{k} \left( \gamma_0 \,U(\xi) + \frac{\gamma_1}{2} \,U(\xi)^2 + \frac{\gamma_2}{3} \,U(\xi)^3 + \frac{\gamma_3}{4} \,U(\xi)^4 \right) + c_1 
    \end{aligned}
\end{equation}

\begin{figure}[!ht]
    \centering
    \subfigure[]{\includegraphics[width=0.47\textwidth]{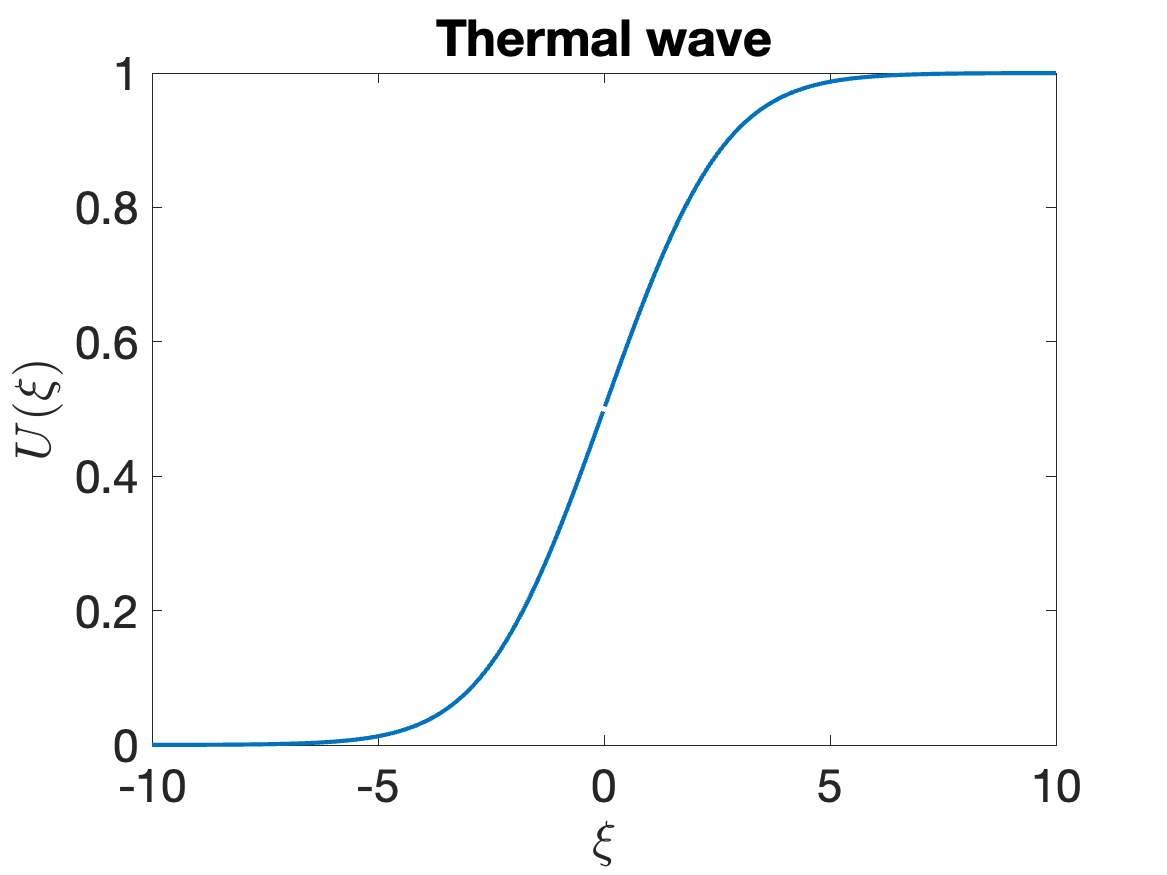}}
    \subfigure[]{\includegraphics[width=0.47\textwidth]{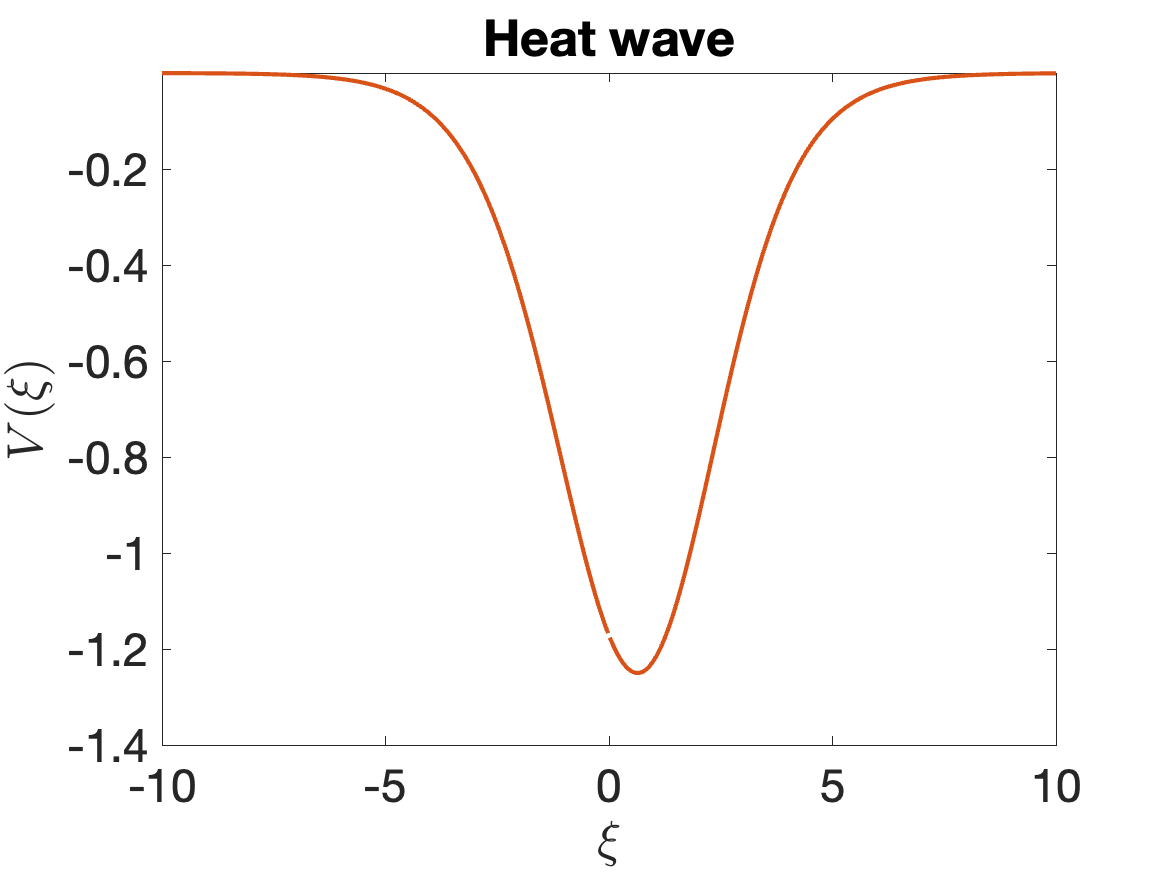}} \\
   \caption{Thermal dark soliton $U(\xi)$ (left) and heat bright soliton $V(\xi)$ (right) solution \eqref{solU_case_n_6_m_3} for $A_0=A_2=1$. Plots are achieved for the following set of parameters: $\alpha=1.25$, $\alpha_0=0.08$, $\alpha=1.25$, $k=0.8$, $w=0.7$, $c_0=1/(w\,\alpha)$, $b_1=-1$, $b_2=1$, $c_1=0$ and $c_2=0$.}
    \label{fig:case_n_6_m_3_ODE}
\end{figure}

\begin{figure}[!ht]
	\centering
    	\subfigure[]{\includegraphics[width=0.45\textwidth]{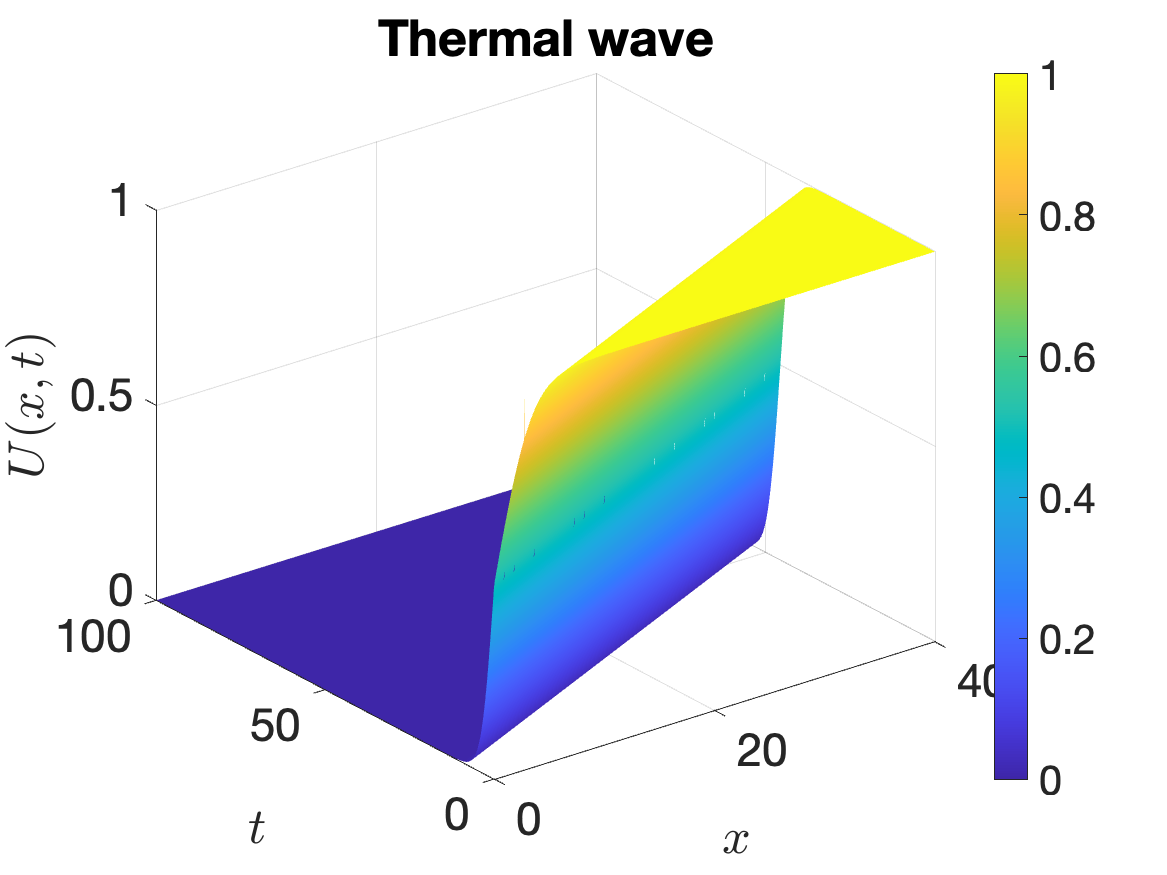}}
    	\subfigure[]{\includegraphics[width=0.45\textwidth]{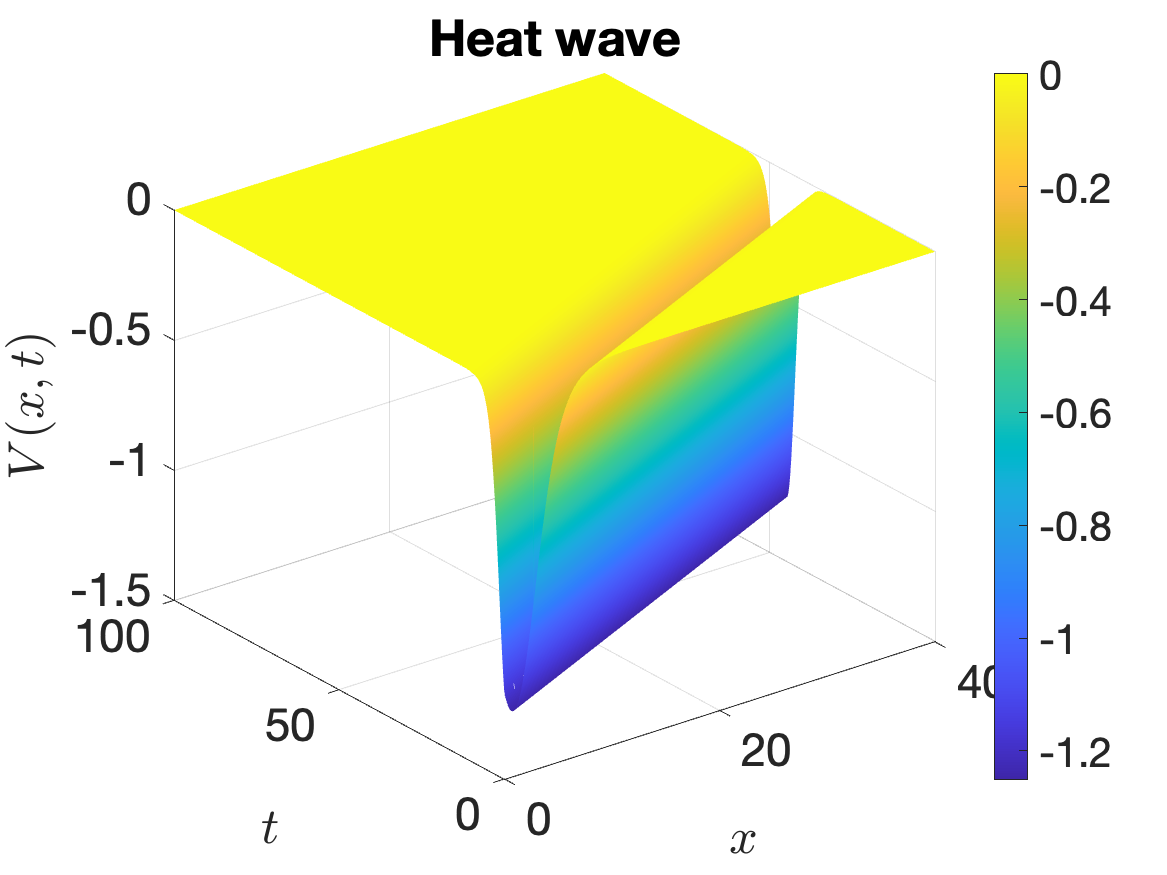}} \\
        \subfigure[]{\includegraphics[width=0.45\textwidth]{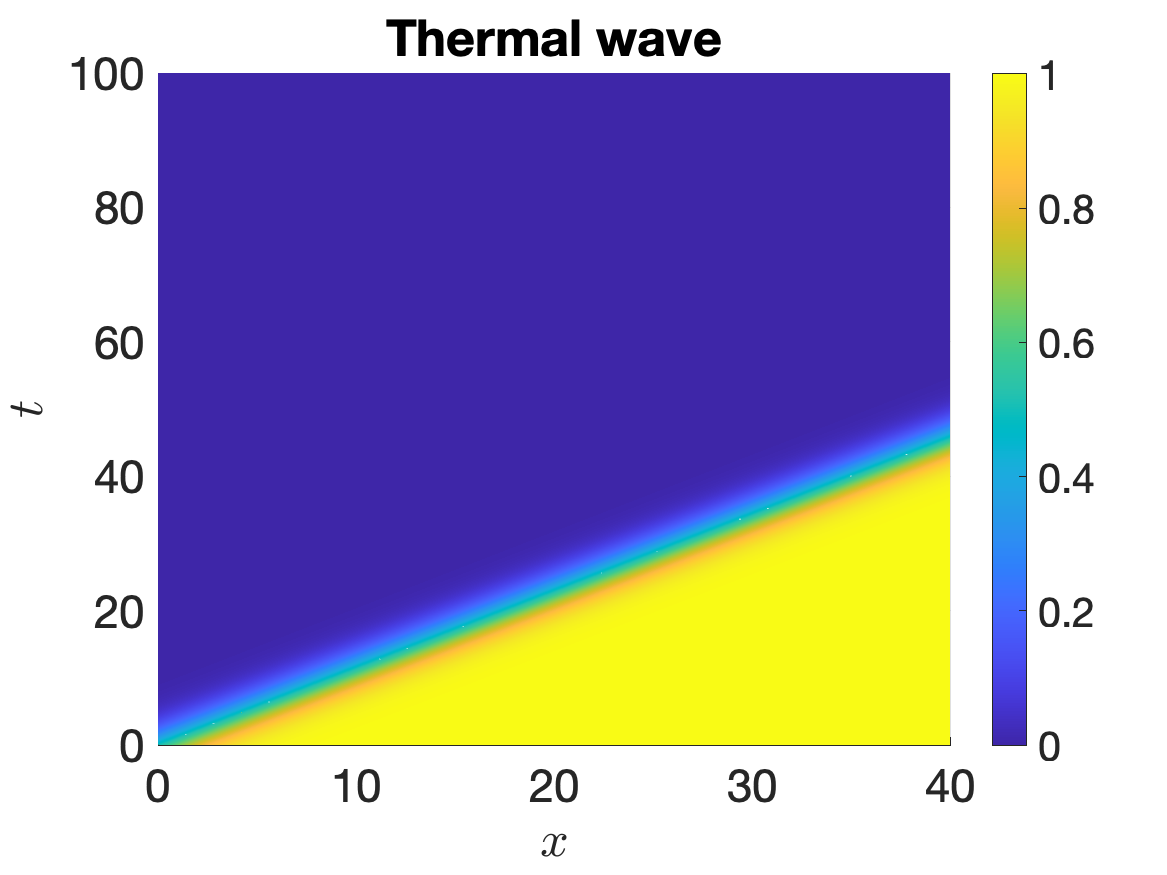}}
    	\subfigure[]{\includegraphics[width=0.45\textwidth]{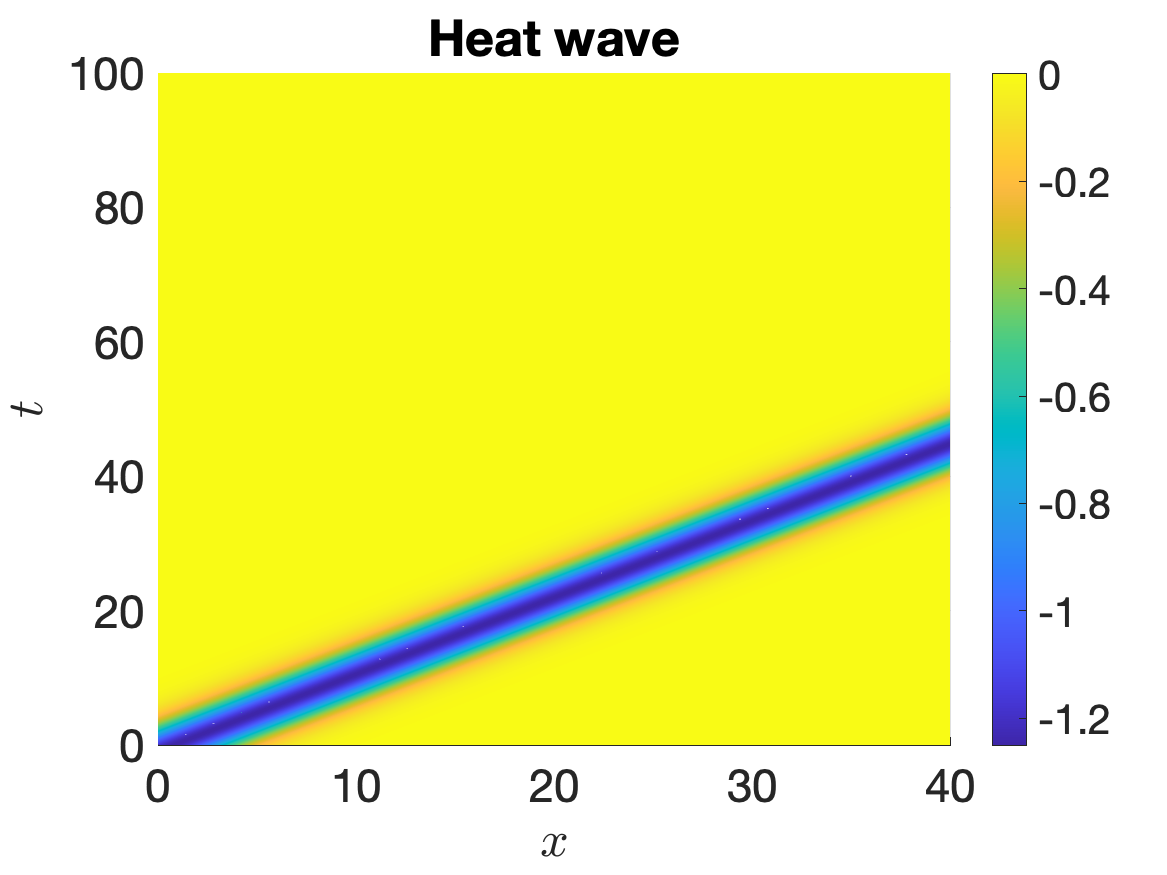}} \\
    	\caption{Plot of the soliton solutions $u(x,t)$ and $v(x,t)$, for $x \in [0,40]$ and $t\in\times[0,100]$, ans $A_0=A_2=1$ with the following parameter values: $\alpha=1.25$, $\alpha_0=0.08$, $\alpha=1.25$, $k=0.8$, $w=0.7$, $c_0=1/(w\,\alpha)$, $b_1=-1$, $b_2=1$, $c_1=0$ and $c_2=0$.}
    	\label{fig:case_n_6_m_3_PDE}
\end{figure}

\begin{figure}[!ht]
    \centering
    \subfigure[]{\includegraphics[width=0.47\textwidth]{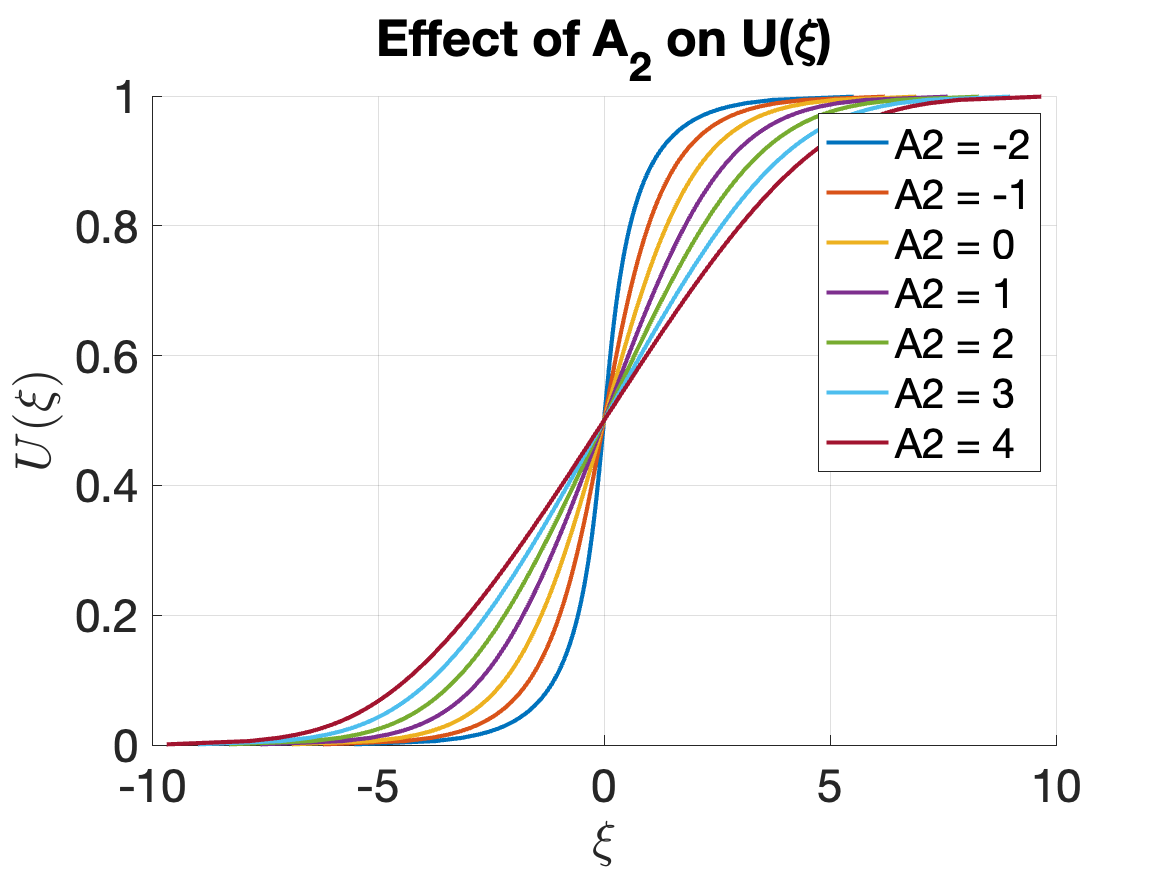}}
    \subfigure[]{\includegraphics[width=0.47\textwidth]{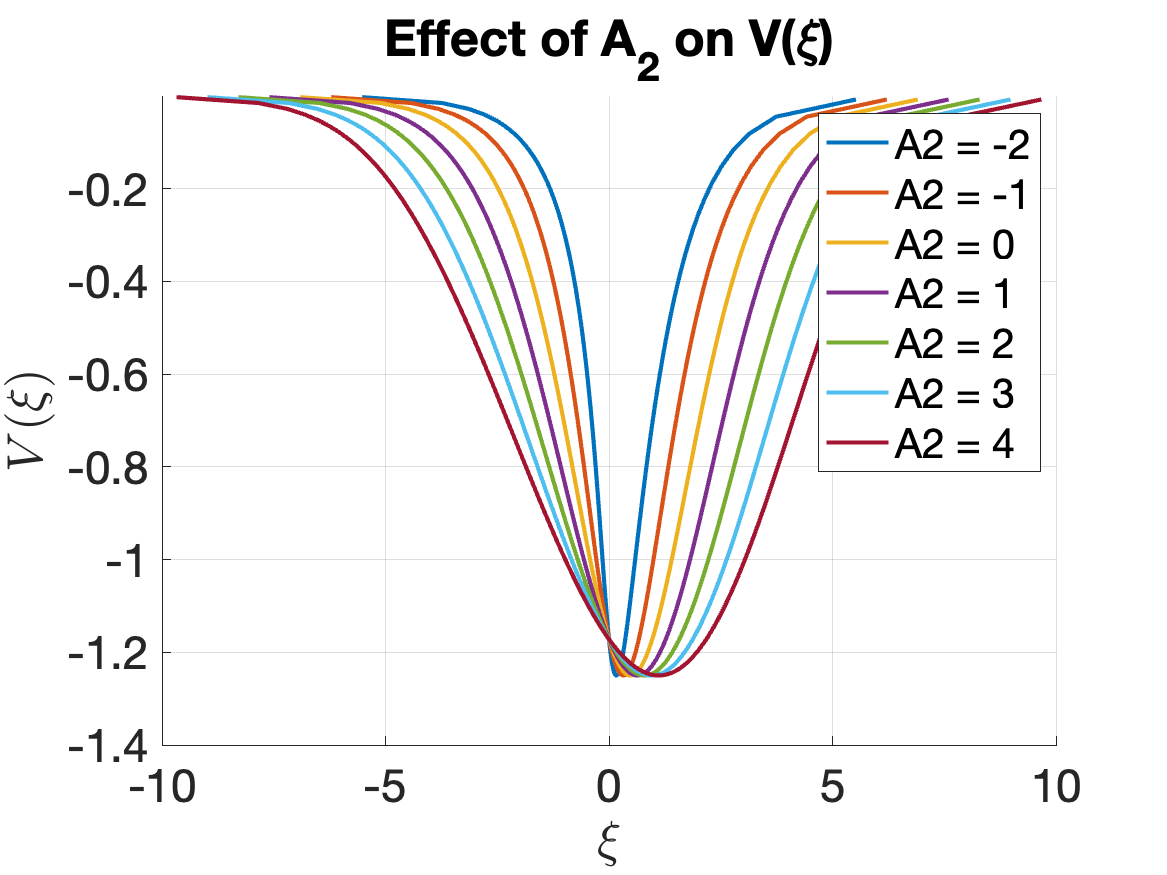}} \\
    \caption{Plots of the dark soliton solutions \eqref{solU_case_n_6_m_3} for $A_0 \neq A_2$, for some values of $A_2$ and fixed $A_0=1$ and the following  parameter values: $\alpha=1.25$, $\alpha_0=0.08$, $\alpha=1.25$, $k=0.8$, $w=0.7$, $c_0=1/(w\,\alpha)$, $b_1=-1$, $b_2=1$, $c_1=0$ and $c_2=0$.}
    \label{fig:case_n_6_m_3_ODE_variation_A2}
\end{figure}

\begin{figure}[!ht]
    \centering
    \subfigure[]{\includegraphics[width=0.47\textwidth]{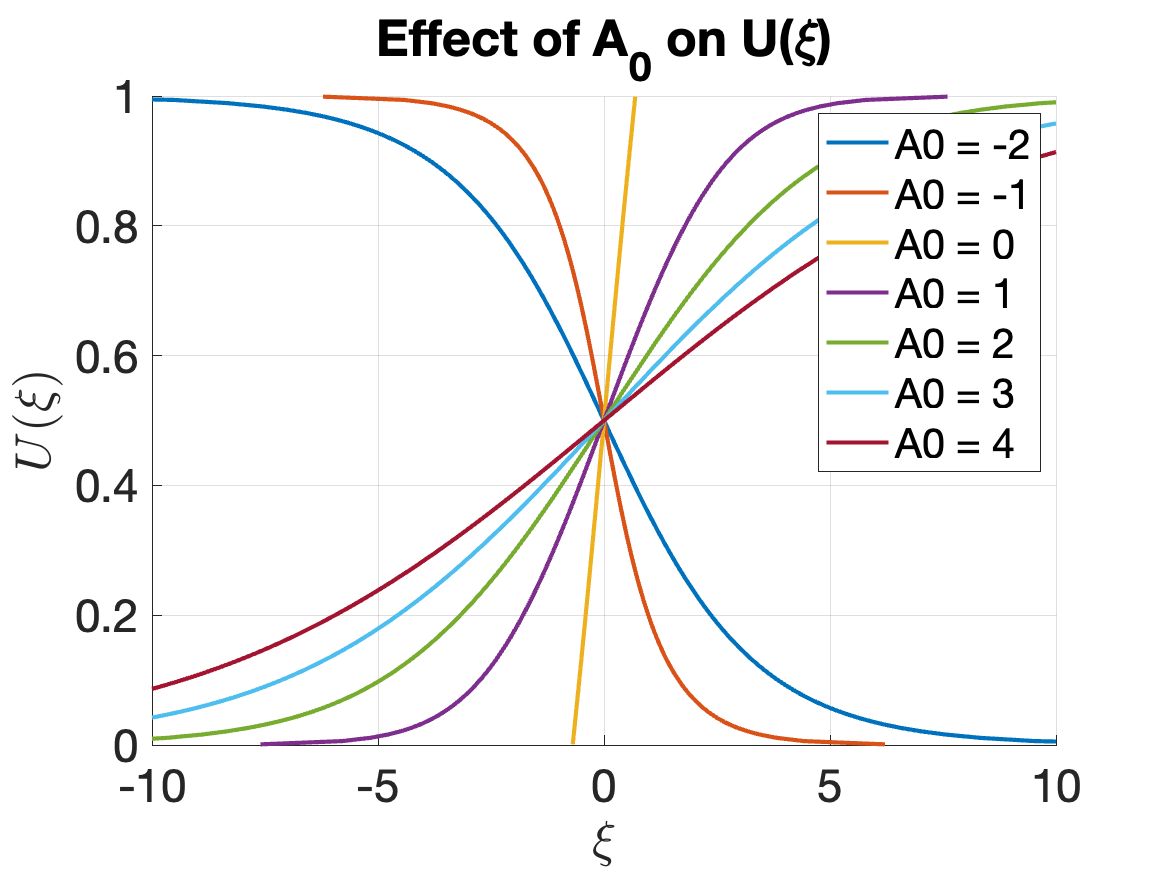}}
    \subfigure[]{\includegraphics[width=0.47\textwidth]{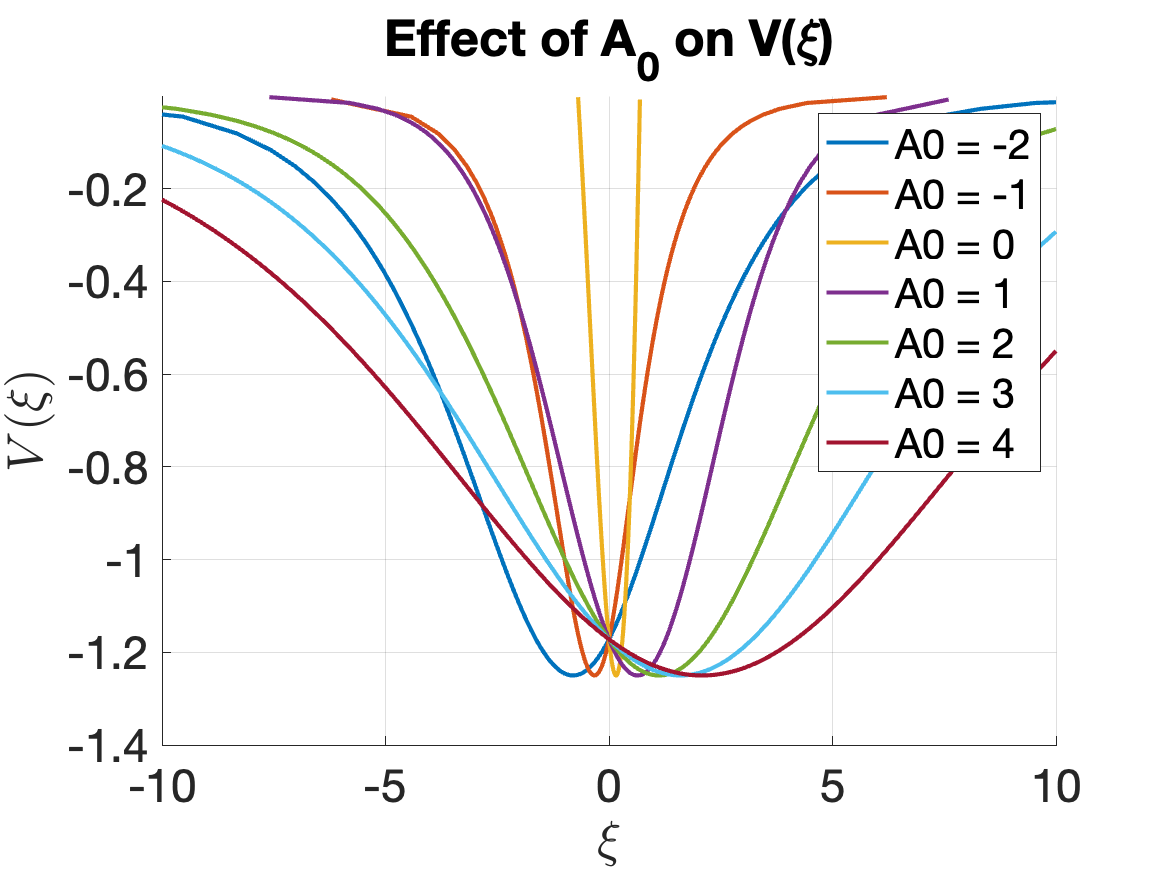}} \\
    \caption{Plots of the dark soliton solutions \eqref{solU_case_n_6_m_3} for $A_0 \neq A_2$, for some values of $A_0$ and fixed $A_2=1$ and the following  parameter values: $\alpha=1.25$, $\alpha_0=0.08$, $\alpha=1.25$, $k=0.8$, $w=0.7$, $c_0=1/(w\,\alpha)$, $b_1=-1$, $b_2=1$, $c_1=0$ and $c_2=0$.}
    \label{fig:case_n_6_m_3_ODE_variation_A0}
\end{figure}

The behaviour of these solutions is plotted in Figures \ref{fig:case_n_6_m_3_ODE} and \ref{fig:case_n_6_m_3_PDE}. Note that they   satisfy the Definition~\ref{definition1} of soliton, in  particular,  $U(\xi)$ is a dark soliton because it shows two asymptotes in $\pm \infty$ (like the $\tanh(\xi)$ function), and  $V(\xi)$ is a bright soliton because it shows the same asymptote in $\pm \infty$ (like the $\operatorname{sech}(\xi)$ function). In Figures \ref{fig:case_n_6_m_3_ODE_variation_A2} and \ref{fig:case_n_6_m_3_ODE_variation_A0} the solutions are depicted as $A_2$ varies with $A_0$ fixed, and as $A_0$ varies with $A_2$ fixed, respectively.

\section{\textsc{Conclusions}}\label{sec6}
The growing interest in nonlinear phenomena led us to consider in this paper a nonlinear version of the Maxwell-Cattaneo-Vernotte equation for the dynamics of the heat flux, where the relaxation time $\tau$, the mass density $\rho$ and the thermal conductivity $\lambda$ are assumed to be smooth functions of the temperature $T$. This paper generalizes the previous mathematical model considered by K\'ov\'acs and Rogolino \cite{kovacs2020MCV1D}, where $\tau$, $\rho$ and $\lambda$ were taken to be linear functions of the temperature $T$. Here, instead, we assume that $\tau(T)$, $\rho(T)$ and $\lambda(T)$ belong to $\mathcal{C}^{\infty}(\mathbb{R})$, and we approximte them through their Taylor expansions around a reference temperature $T_0$, as reported in \eqref{funzioni_lam} and \eqref{funzioni_tau}. 

The mathematical model proposed in this paper is compatible with the constraints imposed by the thermodynamics of irreversible processes. 
The governing equations, obtained by coupling the MCV equation with the internal energy balance, have been reduced to a one-dimensional domain (cylindrical shape) in order to investigate the propagation of thermal and heat waves.
By introducing a moving frame of reference $\xi=kx-wt$, it is possible to follow the wave moving with the velocity $v=w/k$ and to transform the original system of partial differential equations into a system of coupled ordinary differential equations. These ODEs have been analyzed  and we have found all the admissible solutions of the mathematical model.  

Section~\ref{sec5} is devoted to the study of thermal and heat waves in several physically relevant situations, including those considered in Ref.~\cite{kovacs2020MCV1D}, as well as the propagation of one-dimensional single-soliton, and trains of two solitons. These cases are closely connected to the truncation of the Taylor expansions. In particular, the case $n=m=1$ leads to the implicit solution \eqref{sol_U} and the stationary solutions, which may be of interest, for instance, in one-dimensional nanosystems for storing information.
For  $m=1$ and $n=2$, the model admits both thermal and heat soliton solutions, namely dark solitons (tanh-type) for the temperature field and bright solitons (sech-type) for the heat flux.
In the case $n=6$ and $m=3$, the structure  of the solutions leads to the presence of two $\operatorname{artanh}$ terms in the implicit expression of the thermal waves (in contrast to the presence of single $\operatorname{artanh}$ term appearing in the previous subsection). This situation corresponds to the overlapping of two soliton waves travelling  at the same speed (determined by the variable $\xi$), as illustrated in Figs. \ref{fig:case_n_6_m_3_ODE} and \ref{fig:case_n_6_m_3_PDE}, which confirm the presence of other solitons.

This behavior can be further generalized to the case  $n=2m$,
which allows a balance between  the two terms in the numerator of the integral  \eqref{integral_U}, together with the additional requirement that $m$ be odd, so as to obtain an even polynomial in the denominator of the same integral \eqref{integral_U}.
By a suitable choice of the parameters (following the same procedure used in  subsection~\ref{sec53}) a series of $\operatorname{artanh}$ terms  (specifically $(m+1)/2$) emerges upon integration of equation  \eqref{integral_U}. The resulting solution corresponds to the overlapping of $(m+1)/2$ soliton propagating at the same speed $w/k$, which mimics the propagation of a single soliton. These results contribute significantly to the propagation of thermal signal in nanoscale systems, such as nanowires.  In particular, a single soliton can contain a significant amount of information, depending on the parameter $m$, which is related to the number of zeros in the denominator  of the integral \eqref{integral_U}.  This observation highlights the crucial role played by the functional dependence of the relaxation time and thermal conductivity, as captured through their Taylor expansions.

Further investigations will be devoted to the analysis of other significant cases. In particular, a comparison with experimental data in materials exhibiting thermal wave behavior would be of great interest in order to assess the physical applicability of the proposed model. Furthermore, the identification of additional polynomial degrees leading to significant thermal and heat wave solutions represents a natural extension of the present analysis. 

\bigskip
\paragraph*{Acknowledgements} 
The authors acknowledge the support of GNFM–INdAM (Istituto Nazionale di Alta Matematica).

The paper is partially supported by the PNRR project  ECS00000022 "SiciliAn MicronanOTecH Research And Innovation CEnter (SAMOTHRACE)" and the PRIN project 2022TMW2PY 002 “Transport phonema in low dimensional structures: models, simulations and theoretical aspects”

\bibliographystyle{unsrt} 
\bibliography{biblio}

@article{cattaneo1948sullaconduzione,
  title={Sulla conduzione del calore},
  author={C. Cattaneo},
  journal={Atti del Seminario Matematico e Fisico dell'Università di Modena},
  volume={3},
  pages={3--21},
  year={1948}
}

@article{jou2008EIT,
  title={Extended irreversible thermodynamics of heat transport: A brief introduction},
  author={D. Jou and J. Casas-V{\'a}zquez and G. Lebon},
  journal={Proceedings of the Estonian Academy of Sciences},
  volume={57},
  number={3},
  year={2008},
  publisher={Citeseer}
}

@article{toth2025unified,
  title={A unified mixed hp-finite element framework for modeling laser pulse heating processes in the refined thermodynamics},
  author={B. Toth},
  journal={International Journal of Heat and Mass Transfer},
  volume={238},
  pages={126456},
  year={2025},
  publisher={Elsevier}
}

@book{jou2010EIT,
  title={Extended irreversible thermodynamics: non-equilibrium equations of state},
  author={D. Jou and G. Lebon and J. Casas-V{\'a}zquez},
  year={2010},
  publisher={Springer}
}

@article{kapitza1941heat,
  title={Heat transfer and superfluidity of helium II},
  author={P. L. Kapitza},
  journal={Physical Review},
  volume={60},
  number={4},
  pages={354},
  year={1941},
  publisher={APS}
}

@incollection{dresner1984transient,
  title={Transient heat transfer in superfluid helium—part II},
  author={L. Dresner},
  booktitle={Advances in Cryogenic Engineering: Volume 29},
  pages={323--333},
  year={1984},
  publisher={Springer}
}

@phdthesis{mcnelly1974PhD,
  title={Second sound and anharmonic processes in isotopically pure alkali-halides},
  author={T.F. McNelly},
  year={1974},
  school = {Cornell University}
}

@phdthesis{munafo2024PhD,
  title={Non-linear heat transfer beyond Fourier’s Law: analytical and numerical investigations},
  author={C.F. Munaf\`o},
  year={2024},
  school={Universit{\`a} degli Studi di Palermo}
}

@article{walker1971thermalconductivity,
  title={Thermal conductivity, second sound, and phonon-phonon interactions in NaF},
  author={H.E. Jackson and C.T. Walker},
  journal={Physical Review B},
  volume={3},
  number={4},
  pages={1428--1439},
  year={1971},
  publisher={APS}
}

@article{ramos2024numerical,
  title={Numerical analysis of the Maxwell-Cattaneo-Vernotte nonlinear model},
  author={A. Ramos and  A. Campelo and M. Freitas and R.  Kov{\'a}cs},
  journal={Journal of Thermal Stresses},
  volume={47},
  number={9},
  pages={1262--1276},
  year={2024},
  publisher={Taylor \& Francis}
}

@article{munafo2024nonlinear,
  title={Nonlinear thermal analysis of two-dimensional materials with memory},
  author={C.F. Munaf{\`o} and P. Rogolino and R. Kov{\'a}cs},
  journal={International Journal of Heat and Mass Transfer},
  volume={219},
  pages={124847},
  year={2024},
  publisher={Elsevier}
  }

@article{munafo2025comparison,
  title={Comparison of two nonlinear formulations of the Maxwell-Cattaneo equation in heat pulse transmission},
  author={C.F. Munaf{\`o} and P. Rogolino and D. Jou},
  journal={Applied Mathematical Modelling},
  volume={137},
  pages={115684},
  year={2025},
  publisher={Elsevier}
}

@article{atkins1950velocity,
  title={The velocity of second sound below 1 K},
  author={K.R. Atkins and D.V. Osborne},
  journal={The London, Edinburgh, and Dublin Philosophical Magazine and Journal of Science},
  volume={41},
  number={321},
  pages={1078--1081},
  year={1950},
  publisher={Taylor \& Francis}
}

@article{lane1947second,
  title={Second sound in liquid helium II},
  author={C.T. Lane and H. A. Fairbank and W.M. Fairbank},
  journal={Physical Review},
  volume={71},
  number={9},
  pages={600},
  year={1947},
  publisher={APS}
}

@article{kovacs2020MCV1D,
  title={Numerical treatment of nonlinear Fourier and Maxwell-Cattaneo-Vernotte heat transport equations},
  author={R. Kov{\'a}cs and P. Rogolino},
  journal={International Journal of Heat and Mass Transfer},
  volume={150},
  pages={119281},
  year={2020},
  publisher={Elsevier}
}

@article{onsager1931reciprocal,
  title={Reciprocal relations in irreversible processes. I.},
  author={L. Onsager},
  journal={Physical review},
  volume={37},
  number={4},
  pages={405},
  year={1931},
  publisher={APS}
}

@article{joseph1989heat,
  title={Heat waves},
  author={D. Joseph and L. Preziosi},
  journal={Rev. Mod. Phys.},
  volume={61},
  number={},
  pages={41-73},
  year={1989},
  publisher={}
}

@book{muller1998extended,
  title={Rational Extended Thermodynamics},
  author={I. M{\"u}ller and T. Ruggeri},
  year={1998},
  publisher={Springer}
}

@article{rogolino2018generalized,
  title={Generalized heat-transport equations: parabolic and hyperbolic models},
  author={P. Rogolino and R. Kov{\'a}cs and P. V{\'a}n and V. A. Cimmelli},
  journal={Continuum Mechanics and Thermodynamics},
  volume={30},
  number={6},
  pages={1245--1258},
  year={2018},
  publisher={Springer}
}

@article{sciacca2020heat,
  title={Heat solitons and thermal transfer of information along thin wires},
  author={M. Sciacca and  F. X. Alvarez and D. Jou and J. Bafaluy},
  journal={International Journal of Heat and Mass Transfer},
  volume={155},
  pages={119809},
  year={2020},
  publisher={Elsevier}
}

@article{Sciacca_WM_2022thermal,
  title={Thermal solitons in nanotubes},
  author={M. Sciacca and I. Carlomagno and A. Sellitto},
  journal={Wave Motion},
  pages={102967},
  year={2022},
  publisher={Elsevier}
}

@article{Sciacca_JNET47_2022nonlinear,
  title={Nonlinear thermal transport with inertia in thin wires: thermal fronts and steady states},
  author={M. Sciacca and D. Jou},
  journal={Journal of Non-Equilibrium Thermodynamics},
  volume={47},
  number={2},
  pages={187--194},
  year={2022},
  publisher={De Gruyter}
}

@article{Sciacca_PD423_2021two,
  title={Two relaxation times and thermal nonlinear waves along wires with lateral heat exchange},
  author={M. Sciacca},
  journal={Physica D: Nonlinear Phenomena},
  volume={423},
  pages={132912},
  year={2021},
  publisher={Elsevier}
}

@article{Sciacca_JMF62_2021thermal,
  title={Thermal solitons along wires with flux-limited lateral exchange},
  author={M. Sciacca and X.F. Alvarez and D. Jou and J. Bafaluy},
  journal={Journal of Mathematical Physics},
  volume={62},
  number={10},
  year={2021},
  publisher={AIP Publishing}
}

@book{Mongiovi_book2025non,
  title={Non-Equilibrium Thermodynamics of Superfluid Helium and Quantum Turbulence},
  author={M.S. Mongiov{\`\i} and D. Jou and M. Sciacca},
  year={2025},
  publisher={Springer}
}

@book{Agrawal_book2001,
  author={G. P. Agrawal},
  title={Nonlinear {F}iber {O}ptics},
  publisher={Academic Press},
  howpublished={New York},
  year={2001}
}

@BOOK{Hasegawa-book1995,
    AUTHOR={A. Hasegawa and Y. Kodama},
    TITLE={{Solitons in Optical Communications}},
    PUBLISHER={Oxford University Press},
    ADDRESS={New York},
    YEAR={1995}
}

@article{kivshar1998dark,
  title={Dark optical solitons: physics and applications},
  author={Y.S. Kivshar and B. Luther-Davies},
  journal={Physics reports},
  volume={298},
  number={2-3},
  pages={81--197},
  year={1998},
  publisher={Elsevier}
}

@article{Frantzeskakis2010dark,
  title={Dark solitons in atomic Bose--Einstein condensates: from theory to experiments},
  author={D.J. Frantzeskakis},
  journal={Journal of Physics A: Mathematical and Theoretical},
  volume={43},
  number={21},
  pages={213001},
  year={2010}
}

\end{document}